\documentclass{ws-rv975x65}
\usepackage{subfigure}     
\usepackage{ws-rv-van}     
\makeindex

  \newcommand{\myskip}[1]{}

\usepackage[english]{babel}

\myskip{}
{
\usepackage[latin9]{inputenc}
\usepackage{color}
\definecolor{note_fontcolor}{rgb}{0.80078125, 0.80078125, 0.80078125}
\usepackage{amsbsy}
\usepackage{amstext}
\usepackage{lineno}
}

  \newcommand{\mytext}[1]{}

\newcommand{\tr}{{\rm tr}} 

 \newcommand{\scriptD}{{\hat{\cal D}}}
 \newcommand{\scriptN}{{\cal N}}
 \newcommand{\scriptR}{\hat{\cal R}}
\newcommand{\scriptE}{{\cal E}}

 \newcommand{\up}{\uparrow}
  \newcommand{\down}{\downarrow}
 \newcommand{\Up}{\Uparrow}
  \newcommand{\Down}{\Downarrow}
    
  \newcommand{\half}{\frac{1}{2}}
  
  \newcommand{\BEQ }{\begin{eqnarray}}
  \newcommand{\EEQ }{\end{eqnarray}}
  \newcommand{\BEA}{\begin{eqnarray}}
  \newcommand{\EEA}{\end{eqnarray}}
  \newcommand{\nn}{\nonumber}
    \renewcommand{\d}{{\rm d}}

\newcommand{\sub}{{\rm sub}}

\newcommand{\tf}{t_{\rm f}}

  \newcommand{\trunc}{{\rm trunc}}

  \renewcommand{\thesection}{\arabic{section}}
   \renewcommand{\theequation}{\thesection.\arabic{equation}}
     \renewcommand{\thesection}{\arabic{section}}
 \renewcommand{\thesubsection}{\thesection\arabic{subsection}}
    \renewcommand{\theequation}{\thesection\arabic{equation}}
\renewcommand{\thefigure}{\thesection\arabic{figure}}

\makeatletter

 \let\oldforeign@language\foreign@language
 \DeclareRobustCommand{\foreign@language}[1]{%
   \lowercase{\oldforeign@language{#1}}}

\makeatother

\begin{document}

\markboth{Nieuwenhuizen, Perarnau Llobet, Balian}{Dynamical models for quantum measurements}


\setcounter{chapter}{7}

\chapter[Dynamical models for quantum measurements]{Lectures on dynamical models for quantum measurements}

\author
[Nieuwenhuizen, Perarnau-Llobet and Balian]
{Theo M. Nieuwenhuizen$^{(1)}$, Marti Perarnau-Llobet$^{(2)}$ and Roger Balian$^{(3)}$}

\address{
$^{(1)}$ Institute for Theoretical Physics, University of Amsterdam, Science Park 904,  Postbus 94485, 
1090 GL Amsterdam, The Netherlands \\
t.m.nieuwenhuizen@uva.nl} 
\address{
$^{(2)}$ 
ICFO -- The Institute of Photonic Sciences,
Mediterranean Technology Park,
Av. Carl Friedrich Gauss, 3,
08860 Castelldefels (Barcelona), Spain\\
marti.perarnau.llobet@gmail.com}
 \address{$^{(3)}$  Institute de Physique Th\'eorique, CEA Saclay, 91191 Gif-sur-Yvette  cedex, France}




\begin{abstract}
In textbooks, ideal quantum measurements are described in terms of the tested system only by the collapse postulate and Born's rule.
This level of description offers a rather flexible position for the interpretation of quantum mechanics.
Here we analyse an ideal measurement as a process of interaction between the tested system S and an apparatus A, so as to derive the properties 
postulated in textbooks. We thus consider within standard quantum mechanics
 the measurement of a quantum spin component $\hat s_z$ by an apparatus A, being a magnet coupled to a bath.
 We first consider the evolution of the density operator of S+A describing a large set of runs of the measurement process. 
 The approach describes the  disappearance of the off-diagonal terms (``truncation'')  of the density matrix as a physical
effect due to A, while the registration of the outcome has classical features due to the large size of the pointer variable,
the magnetisation. A quantum ambiguity implies that the density matrix at the final time can be decomposed 
on many bases, not only the one of the measurement. 
This quantum oddity prevents to connect individual outcomes to measurements, a difficulty known as the ``measurement problem''.
It is shown that it is circumvented by the apparatus as well, since the evolution in a small time interval erases all decompositions, except the 
one on the measurement basis. 
Once one can derive the outcome of individual events from quantum theory, the so-called ``collapse of the wave function'' or the ``reduction of the state'' 
appears as the result of a selection of runs among the original large set. 
Hence nothing more than standard quantum mechanics is needed to explain features of measurements.
The employed statistical formulation is advocated for the teaching of quantum theory.
\end{abstract}

 {Keywords: dynamics of quantum measurements; 
quantum measurement 

problem; 
 ensemble   interpretation}

PACS:  87.16.Nn, 05.40.-a, 05.60.-k


\body

\renewcommand{\thesection}{\arabic{section}}
\section{Introduction}
\setcounter{equation}{0}\setcounter{figure}{0}
\renewcommand{\thesection}{\arabic{section}.}
\label{section.3}

Quantum mechanics is our most fundamental theory at the microscopic level, and its successes are innumerable, see, e. g., Ref. \citen{FQMT11}. 
However, although one century has passed 
since its beginnings, its interpretation is still subject to discussions. What is the status of wave functions facing reality? 
Are they just a tool for making predictions \cite{deMuynck}, or do they describe individual objects? How should we understand strange features such as Bell's inequalities? 
To answer such questions, we have to elucidate the only point of contact between theory and reality, to wit, measurements. 
Thus, a proper understanding of quantum measurements may provide useful lessons for a sensible interpretation of quantum theory, 
lessons not learnable from a ``black box'' approach where only the measurement postulates are employed.

A measurement should be analyzed as a dynamical process in which the tested quantum system S interacts with another quantum system, the apparatus A.
This apparatus reaches at the end of the process one among 
several possible configurations. They are characterized by the indication of a pointer, that is, by the value of a {\it pointer variable} of A which we can observe or register, 
and which provides us with information about the initial state of S. This transfer of information from S to A, allowed by the coupling between S and A, thus involves 
a perturbation of A. Moreover, in quantum mechanics, the interaction process also modifies S in general; this is understandable since the apparatus is much larger 
than the system.\footnote{When we speak about ``the system'', we always mean: an ensemble of identically prepared systems, and for ``the measurement''
an ensemble of measurements performed on the ensemble of systems. As in classical thermodynamics, the ensemble can be real or Gedanken.}

For conceptual purposes, it is traditional to consider {\it ideal measurements}, although these can rarely be performed in actual experiments. 
Ideal measurements are those which produce the weakest possible modification of S. 
In textbooks, ideal quantum measurements are usually treated, without caring much about the apparatus, by postulating two properties about the fate 
of the tested system. {\it Born's rule} provides the probability of finding the eigenvalue $s_i$ of the observable $\hat s$ which is being measured. The resulting final 
state of S is expressed by von Neumann's {\it collapse}; it is obtained by projecting the initial state over the eigenspace of $\hat s$ associated with $s_i$. 
Clearly, there is a gap with the practice of reading off the pointer variable of a macroscopic apparatus in a laboratory.  Moreover, it is not satisfactory to complement 
the principles of quantum mechanics with such ``postulates''.  In a laboratory, the apparatus itself is a quantum system coupled to S, and a measurement
is a dynamical process involving S+A, so that one deals with two coupled quantum systems and therefore hopes to be able, without introducing new postulates, 
to describe the evolution of the coupled system S+A and its outcome by just solving its quantum  equations of motion. 
The above properties of ideal measurements will then appear not as postulates but as mere consequences of quantum theory applied  to the system S+A. 
 Dynamical models for measurements have therefore been studied, with various benefits. The literature on this subject 
has been reviewed in ref. \citen{ABNOpus}. In particular, a rich enough but still tractable model has been introduced a decade ago, the Curie--Weiss model for the measurement of 
the $z$--component of a spin $\half$  by an apparatus that itself consists of a piece of matter containing many such spins coupled to a thermal phonon bath \cite{ABNCW}. 

In most of such models, the apparatus is a {\it macroscopic object} having several stable states, each of which is characterised by some value of the pointer variable. 
Its initial state is metastable; by itself it would go after a very long time to one among these stable states. In the presence of a coupling with S, such a transition is 
triggered by the measurement process, in such a way that the eigenvalues $s_i$ of $\hat s$ and the indications $A_i$ of the pointer
become fully correlated and can be read off -- the two purposes of the measurement. 

In an ideal measurement, the tested observable $\hat s$ commutes with the Hamiltonian, implying that in the diagonal basis $ \{ |i \rangle \} $ of $\hat s$ the various 
sectors of the density matrix remain decoupled during the whole measurement. 
They will thus evolve independently, driven by different aspects of the physics. The off-diagonal blocks of the density 
matrix of S+A are the ones for which S is described by $|i\rangle \langle j|$ with $j\neq i$; they are sometimes called ``Schr\"odinger cat terms''. 
In the considered models, they evolve due to a dephasing mechanism known from NMR (MRI) physics and/or due to a decoherence
 mechanism produced by a coupling of the pointer with a thermal bath.  As a consequence, the effects of these off-diagonal blocks disappear in incoherent sums of
 phase factors, so for all practical purposes they can be considered as  tending to zero (see, e. g., ref.  \citen{vanKampen}), 
 because their  contributions 
 to the state of S are suppressed. As for each of the diagonal blocks $|i\rangle \langle i|$, its evolution describes the phase transition of the apparatus 
 from its initial metastable state to its stable state correlated with the measured eigenvalues $s_i$;
  this process, which involves a decrease of free energy, requires a dumping of energy in a bath.
The time scales of the two processes are different: the truncation happens rather fast; it involves no energy transfer and has resemblance to the ${\cal T}_2$ time
of NMR physics, while the registration does involve energy transfer to the bath, with resemblance to the ${\cal T}_1$ process on its longer time scale.

If we regard the bath, introduced in most models, as being part of the apparatus, we can treat S+A as an {\it isolated system}. If we were dealing with pure states, 
 its dynamics would  be governed by the Schr\"odinger equation. However, the apparatus being macroscopic, we have to resort to
  {\it quantum statistical mechanics} \cite{Balian_book}. 
 We therefore rely on a formulation of quantum mechanics, recalled below in section 1.1, which encompasses ordinary quantum mechanics but is also adapted to 
 describe macroscopic systems, for instance in solid state physics, in the same way as classical statistical mechanics is adapted to describe large classical systems. 
 The state of S+A is therefore not a pure state, but a statistical mixture.  Wave vectors for A are thus replaced by density operators, describing mixed states.
As S+A is an isolated system, the evolution of its state (i.e., its density operator) is governed by the Liouville--von Neumann equation,
which replaces the  Schr\"odinger equation. We then run into the {\it irreversibility paradox}. Both above-mentioned evolutions, 
 diagonal and off-diagonal, are obviously irreversible, whereas the Liouville--von Neumann evolution is {\it unitary and therefore reversible}. 
 Then, how can this equation give rise to an increase of entropy for S+A?
 As usual, we will solve below this paradox more or less implicitly, by relying on standard methods of statistical mechanics. In particular, 
  acknowledging that our interest lies only in properties that can be observed  on practical timescales,
 we are allowed to discard correlations between a macroscopic number of degrees of freedom; 
 we are also allowed to forget about recurrences that would occur after a very large recurrence time.

\subsection{Outline    }
 
The present course focuses on the {\it Curie--Weiss model}, presented in section 2 below and already studied together with its extensions in ref. \citen{ABNOpus}. But the latter 
 article is too detailed for a pedagogical access. We will therefore restrict ourselves to a simplified presentation. By accounting for the dynamics of the process for 
 S+A in the framework of quantum statistical mechanics, we wish to explain for this model, within the most standard quantum theory, all the features currently 
 attributed to ideal measurements. 
 
 Such features arise due to the physical interaction between S and A, and they are independent of the different interpretations of quantum mechanics.
 The state of the compound system S+A is therefore represented by a time-dependent density operator $\scriptD$ which evolves according to the 
 Liouville--von Neumann equation. At the initial time, it is the product of the state $\hat r(0)$ of S that we wish to test, by the metastable state $\scriptR(0)$ of A 
 prepared beforehand and ready to evolve towards a stable state\footnote{The initial metastable state realises a ``ready'' state of the pointer, 
 ``ready'' to give an indication when a measurement is performed.
Metastability occurs typically in apparatuses, for example in photo multipliers and in our retina. Through its phase transition towards
a stable state, it allows a macroscopic registration of a microscopic quantum signal.}. 
  While $\scriptD(t)$ encompasses our whole information about S+A, it is a mathematical object, 
 the interpretation of which will only emerge at the end of the measurement process, since we can reach insight about the reality of S only through observation 
 of the outcomes (see Section 6 below). 

It is important to realize that pure states, or wave functions, are not proper descriptions of macroscopic systems.\footnote{One of the present authors 
has termed ``the right of every system to have its own wave function'' the ``fallacy of democracy in Hilbert space''.}
Quantum mechanics deals with our {\it information} about systems, which can be coded only in density operators 
representing statistical mixtures.\footnote{Hence the ``collapse postulate'': after the measurement we can update our information about the system.} 
Although it is our most precise theory, it does not deal with properties of individual systems, and thus has a status comparable to statistical  classical mechanics.
In a measurement, the apparatus is macroscopic and measurement theories cannot rely on pure states. 
In the statistical formulation employed in the present paper, this is regarded as unphysical, because
 only a few degrees of freedom for the ensemble of systems can be controlled in practice, 
 so that only ensembles of small systems can be in a pure state. 
 Nevertheless, one encounters many pure-state discussions of measurement in the literature, 
in particular when it is postulated that the apparatus is initially in a pure state. 
Likewise, it is absurd to assume that a cat, also when termed ``Schr\"odinger cat'', can be described by a pure state, being `in a quantum superposition of alive and 
dead''.\footnote{Indeed, one can never have so much information that the Gedanken ensemble of cats may be described by a pure state.}

In this {\it statistical formulation} of quantum mechanics, advocated in ref. \citen{ABNOpus}, a density operator, or ``state'' $\scriptD$ presents an analogy with a standard probability 
 distribution, but it has a specifically quantum feature: It is represented by a matrix rather than by a measure over ordinary random variables. The random physical 
 quantities $\hat O$, or observables, are also represented by matrices, and quantities like Tr $\scriptD \hat O$ will come out as expectation values in experiments. 
 Thus, as the ordinary probability theory and the classical statistical mechanics, 
 quantum theory in its statistical formulation does not deal with individual events (see hereto e. g. Ref. \citen{MichJindeRaedtFQMT11}), but with {\it statistical ensembles} of events. 
 The state $\scriptD(t)$ of S+A which evolves during the measurement process describes only a generic situation. If we wish to think of a single measurement, 
 we should regard it as a sample among a large set of runs, all prepared under the same conditions. A problem then arises because, contrary to ordinary probability 
 theory, quantum mechanics is irreducibly probabilistic due to the non-commutative nature of the observables. After having determined the density operator of the 
 ensemble at the final time but without other information, we {\it cannot make statements about individual measurements}. In particular, this knowledge is not 
 sufficient to explain the observation that each run of a measurement yields a unique answer, the so-called {\it measurement problem},
which has remained unsolved till recently.

Anyhow, a first task is necessary, solving the above-mentioned equations of motion, so as to show that standard quantum statistical mechanics is sufficient 
 to provide the outcome $\scriptD(\tf)$ expected for ideal measurements. These equations are written in Section 4, and their solution is worked out in 
 Sections 4 and 5 for the off-diagonal and diagonal blocks of $\scriptD$, respectively.

At this stage, we shall have determined the state $\scriptD(\tf)$ of S+A which accounts for the {\it whole set of runs} of the measurement, and which involves the 
 expected correlations between the tested eigenvalues of $\hat s$ and the indications $A_i$ of the apparatus. We will exhibit in Section 6 the difficulty that 
 prevents us from inferring properties of individual runs from this mixed state. To overcome this difficulty without going beyond quantum theory, we will 
 consider {\it subensembles of runs}, which can still be studied within standard quantum theory. If we are able to select a subensemble characterised by 
 outcomes corresponding to a given 
 value of the pointer, we expect to be able to {\it update our knowledge}, and hence to describe the selected population of compound systems S+A by a 
 {\it new density operator}. The possibility of performing such a selection is a subtle question, which we tackle through considerations about {\it dynamical stability}. 
 We will thus give an idea of a solution of the long standing measurement problem.
 
 The solution of the model thus relies on several steps. First, the density matrix of S+A associated with the full ensemble of runs is truncated, to wit, 
 it loses its off-diagonal blocks (Section 4). Then, its diagonal blocks relax to equilibrium, thus allowing registration into the apparatus of the information 
 included in the diagonal elements of the initial density matrix of S (Section 5). Next we show that a special type of relaxation yields the needed result 
 for the density operator of any subensemble of runs of the measurement (Sections 6.2 and 6.3). 
 Finally, the structure of these density operators affords a natural interpretation of the process for individual runs in spite of quantum difficulties (Section 6.4).

The Curie--Weiss model is sufficiently simple so as to allow interesting generalisations. In Section 7, we present a model which involves two apparatuses 
 that attempt to {\it measure two non-commuting observables}, namely the components of the spin on two different directions. We shall see that, although this 
 measurement is not ideal and although it seems to involve two incompatible observables, performing a large number of runs can provide statistical information on both.  

\renewcommand{\thesection}{\arabic{section}}
\section{A Curie\,--Weiss model for quantum measurements}
\setcounter{equation}{0}\setcounter{figure}{0}
\renewcommand{\thesection}{\arabic{section}.}
\label{section.3}

In this section we give a detailed description of the Curie--Weiss model for a quantum measurement, which was introduced a decade ago~\cite{ABNCW}.
We take for S, the system to be measured, the simplest quantum system, namely a spin $\half$. 
The observable to be measured is its third Pauli matrix $\hat s_z$ = diag($1,-1$), with eigenvalues $s_i$ equal to $\pm1$. 
For an ideal measurement we assume that $\hat s_z$ commutes with the Hamiltonian of S + A.
This ensures that the statistics of the measured observable are preserved in time, a necessary condition to satisfy Born rule.

We take as  apparatus ${\rm A}={\rm M}+{\rm B}$, a model that
simulates a \textit{magnetic dot}: The magnetic degrees of freedom ${\rm M}$
consist of $N\gg1$ spins with Pauli operators $\hat{\sigma}_{a}^{\left(
n\right)  }$  ($n=1,2,\cdots,N$; $a=x$, $y$, $z$), which read for each $n$

\begin{figure}\label{figLoic1}
\centerline{ \includegraphics[width=8cm]{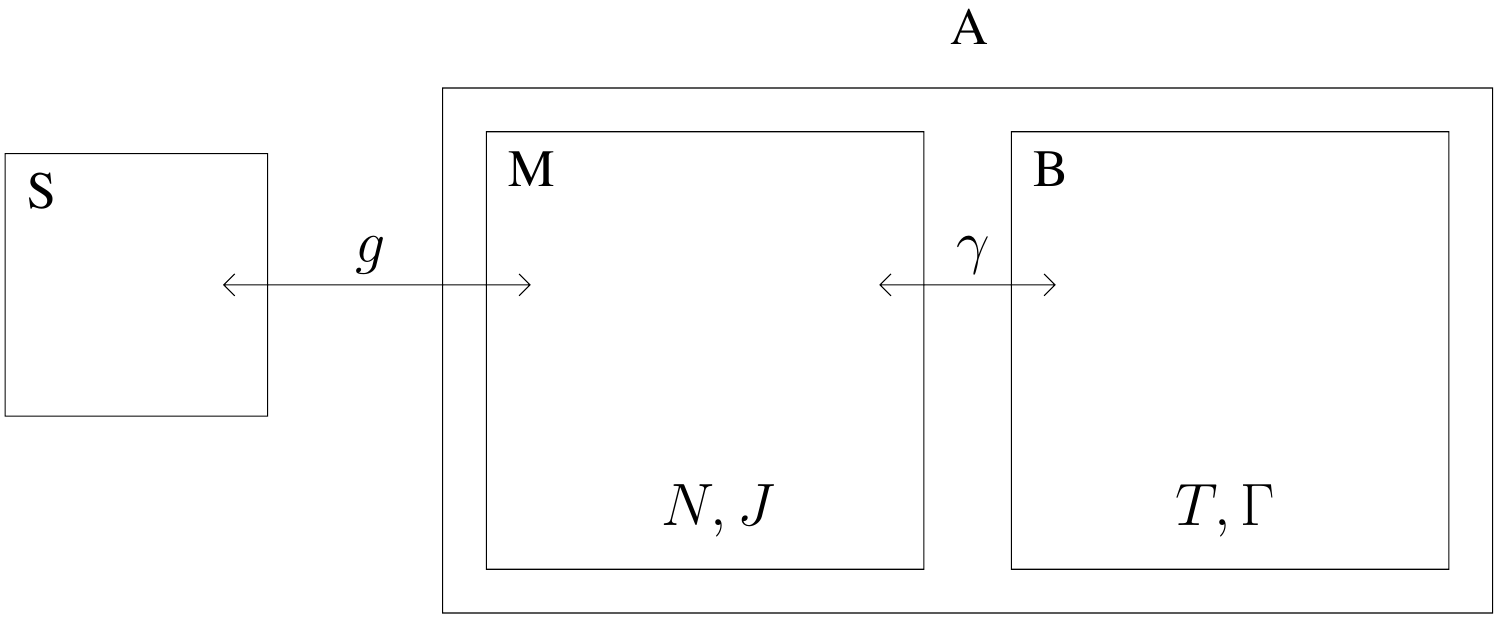}}
\caption{The first version of the Curie-Weiss measurement model and its parameters. The system S is a spin-$\half$ $\hat{\bf s}$.
The apparatus A includes a magnet M and a bath B. The magnet, which acts as a pointer, consists
of $N$ spins-$\half$ coupled to one another through an Ising interaction $J$. The phonon bath B is characterized
by its temperature $T$ and a Debye cutoff $\Gamma$. It interacts with M through a spin-boson coupling 
$\gamma$. The process is triggered by the interaction $g$ between the measured observable $\hat s_z$
and the pointer variable, the magnetization per spin, $\hat m$, of the pointer. 
 To consider the measurement problem, certain weak terms will be added later within the apparatus.  }
\end{figure}

\BEQ \label{sigmaxyz0}
\hat\sigma_x=\left( \begin{array}{ccc@{\ }r} 0 &1\\ 1&0 \end{array} \right),\qquad 
\hat\sigma_y=\left( \begin{array}{ccc@{\ }r} 0 &-i\\ i&0 \end{array} \right),\qquad 
\hat\sigma_z=\left( \begin{array}{ccc@{\ }r} 1 &0\\ 0&-1 \end{array} \right),\qquad 
\hat\sigma_0=\left( \begin{array}{ccc@{\ }r} 1 &0\\ 0&1 \end{array} \right),\qquad 
\EEQ
where $\hat\sigma_0$ is the corresponding identity matrix; 
 {\boldmath{$\hat \sigma$}} $=(\hat\sigma_x,\hat\sigma_y,\hat\sigma_z)$ denotes the vector spin operator.
The non-magnetic degrees of freedom such as phonons behave as a thermal bath ${\rm B}$ (Fig. \ref{figLoic1}). 
As pointer variable we take the
order parameter, which is the magnetization in the $z$-direction (within normalization), as represented by the quantum observable
\begin{equation}
\mytext{\textcurrency hatm=\textcurrency \qquad}
\hat{m}=\frac{1}{N}\sum
_{n=1}^{N}\hat{\sigma}_{z}^{\left(  n\right)  }{ .} \label{hatm=}
\end{equation}
We let $N$ remain finite, which will allow us to keep control of the equations 
of motion. It should however be sufficiently large so as to ensure the existence of thermal equilibrium states with well defined 
magnetization (i.e., fluctuations of the order of $1/\sqrt{N}$).  
At the end of the measurement, the value of the magnetization (either positive or negative) is linked to the two possible outcomes of the measurement, $s_i=\pm 1$.

\subsection{The Hamiltonian}
\label{section.3.2}
We consider the tested system S and the apparatus A as two quantum systems, that are coupled at time $t=0$ and decoupled at time $\tf$.
The full Hamiltonian can be decomposed into terms associated with the system,
with the apparatus and with their coupling:
\begin{equation}
\mytext{\textcurrency ham\textcurrency \qquad}
\hat{H}=\hat{H}_{{\rm S}}
+\hat{H}_{{\rm SA}}+\hat{H}_{{\rm A}}{ .} \label{ham}
\end{equation}

Textbooks treat measurements as instantaneous, which is an idealization. If they are at least very fast, the tested system will hardly undergo dynamics by its own,
so the tested quantity $\hat s$ is practically constant. For an ideal measurement the observable $\hat{s}$ should not proceed at all, so it should commute with $\hat{H}$.
The simplest self-Hamiltonian that ensures this property (no evolution of S without coupling to A), is a constant field $-b_z \hat s_z$, 
which is for our aims equivalent to the trivial case $\hat H_{\rm S}=0$, so we consider the latter.

We take as coupling between the tested system and the apparatus,
\begin{equation}
\mytext{\textcurrency hamham\textcurrency \qquad}
\hat{H}_{{\rm SA}}=-g\hat
{s}_{z}\sum_{n=1}^{N}\hat{\sigma}_{z}^{\left(  n\right)  }=-Ng\hat{s}_{z}
\hat{m}{ .} \label{HSA}
\end{equation}
It has the usual form of a spin-spin coupling in the $z$-direction, and the constant $g>0$ characterizes its strength. As wished, it commutes with
$\hat{s}_{z}$.

The apparatus ${\rm A}$ consists, as indicated above, of a magnet ${\rm M}$ and a phonon bath ${\rm B}$ 
(Fig. \ref{figLoic2}), and its Hamiltonian can be decomposed into

\begin{equation}
\hat{H}_{{\rm A}}=\hat
{H}_{{\rm M}}+\hat{H}_{{\rm B}}+\hat{H}_{{\rm M}{\rm B}}{ .}
\label{HA}
\end{equation}
The magnetic part is chosen as 

\begin{equation}
\hat{H}_{{\rm M}}=-\frac{J}{4} \hat m^4
{ ,} \label{HM=}
\end{equation}
where the magnetization operator $\hat{m}$ was defined by (\ref{hatm=}). It couples all spins {\boldmath{$\hat \sigma$}}$^{(n)}$
symmetrically and anistropically, with the same coupling constant $J$. This Hamiltonian is used to describe superexchange 
interactions in metamagnets.

As we will show in subsequent sections, the Hamiltonian (\ref{HM=}) of M, when coupled to a thermal bath at sufficiently low temperature $T$, 
leads to three locally thermal states for M: a metastable (paramagnetic) state $\hat{\cal R}(0)$ and two stable (ferromagnetic) states, 
$\hat{\cal R}_\Up$ and $\hat{\cal R}_\Down$. A first order transition can then occur from $\hat{\cal R}(0)$  to one of the more stable ferromagnetic states
(for a more realistic set up including first and second order transition we refer the reader to ref. \citen{ABNOpus}).  
An advantage of a first-order transition is the local stability of the paramagnetic state, even below the transition temperature, which ensures a large lifetime. 
It is only by the measurement,  i. e., by coupling to the tested spin, that a fast transition to one of the stable states is triggered. 
This is well suited for a measurement process, which requires the lifetime of the initial state of the apparatus to be larger than the overall measurement time.

The Hamiltonian of the phonon bath, $H_{\rm M} +H_{\rm MB}$, is described in full detail in the Appendix A. The bath plays a crucial role in the Curie-Weiss model,
as it induces thermalization in the states of M. 
{Nevertheless, the degrees of freedom of the bath will be traced out as we are not interested in their specific evolution (recall that the magnetization is the pointer variable)}. 
This induces a non-unitary evolution into the subspace of S+M arising from the unitary evolution of the whole closed system. 

If we assume a very large bath weakly coupled to M, then all the relevant information is compressed in the spectrum of the bath, which
we choose to be quasi-Ohmic:\cite{caldeira,petr,Weiss,Gardiner}

\begin{equation}
\mytext{\textcurrency K\symbol{94}tilde\textcurrency \qquad}
\tilde{K}\left(
\omega\right)  =\frac{\hbar^{2}}{4}\frac{\omega e^{-\left\vert \omega
\right\vert /\Gamma}}{e^{\beta\hbar\omega}-1}{\rm  .} \label{Ktilde}
\end{equation}
where $\beta=1/k_BT$ is the inverse temperature of the bath, the dimensionless paramenter $\gamma$ is the strength of the interaction; and
$\Gamma$ is the  Debye cutoff, which characterizes the largest frequencies of
the bath, and is assumed to be larger than all other frequencies entering our problem.

The spin-boson coupling (\ref{ham4}) between M and B will be sufficient for our purpose up to section 6. This interaction, 
of the so-called Glauber type, does not commute with  $\hat H_{\rm M}$, a property needed for registration, 
since M has to release energy when relaxing from its initial metastable paramagnetic state
(having $\langle\hat m\rangle=0$) to one of its final 
stable ferromagnetic states at the temperature $T$ (having $\langle\hat m\rangle=\pm m_{\rm F}$). 
However, the complete solution of the measurement problem presented in section 6  will require more
 complicated interactions. We will therefore later add a small but random coupling between the spins of M, and in subsection 6.3 a more realistic 
 small coupling  between M and B, of the Suzuki type 
 (that is to say, having terms $\hat\sigma_x^{(n)}\hat\sigma_x^{(n')}+\hat\sigma_y^{(n)}\hat\sigma_y^{(n')}=
 \half (\hat\sigma_+^{(n)}\hat\sigma_-^{(n')}+\hat\sigma_-^{(n)}\hat\sigma_+^{(n')})$, 
 where $\hat\sigma_\pm ^{(n)}=\hat\sigma_x^{(n)}\pm i\hat\sigma_y^{(n)}$),
 which produces flip-flops of the spins of M,  without changing the values of 
 magnetisation and the energy that M  would have with only the terms of (\ref{HM=}).

\subsection{Structure of the states}

\label{section.3.3}

\subsubsection{Notations}

\label{section.3.3.1}
 
\begin{figure}
 \label{figLoic2}
\centerline{ \includegraphics[width=4cm]{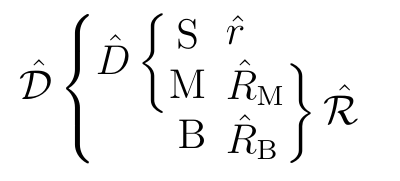}
\hspace{2cm} 
 \includegraphics[width=5.5cm]{FigOpMag1}}
\caption{
 Notations for the density operators of the system S + A and the subsystems M and B of A.
The full density matrix $\hat {\cal D}$ is parametrized by its submatrices $\hat{\cal R}_{ij}$ (with $i,j=\pm1$ or $\uparrow,\downarrow$),
the density matrix $\hat D$ of S + M by its submatrices $\hat R_{ij}$.  The marginal density operator of S is denoted as $\hat r$
and the one of A as $\hat {\cal R}$. The marginal density operator of M itself is denoted as $\hat R_{\rm M}$ and the one of B as $\hat R_{\rm B}$.}
\end{figure}

 Our complete system consists of S+A, that is, S+M+B.
The full state ${\hat{{\cal D}}}$ of the system evolves according to the
Liouville--von~Neumann equation

  \begin{equation}  
  \label{dD}
  i\hbar\frac{\d\hat{\cal D}}{\d t}=[\hat H,\hat{\cal D}]
 \equiv \hat H\hat{\cal D}-\hat{\cal D}\hat H,
    \end{equation} 
which we have to solve. It will
be convenient to define through partial traces, at any instant $t$, the
following marginal density operators: 
$\hat{r}$ for the tested system ${\rm S}$, $\hat{{\cal R}}$ for the apparatus 
${\rm A}$, $\hat{R}_{\rm M}$ for the magnet ${\rm M}$, $\hat{R}_{{\rm B}}$ for the
bath, and $\hat{D}$ for ${\rm S}+{\rm M}$ after elimination of the bath 
(as depicted schematically in Fig. \ref{figLoic2}), according to

\BEA 
&& \hat{r}={\rm {\rm tr}}_{{\rm A}}{\hat{{\cal D}}}{,\qquad}
\hat{{\cal R}} ={\rm {\rm tr}}_{{\rm S}}{\hat{{\cal D}}}{,\qquad} \nn\\&&
\hat{R}_{\rm M}={\rm tr}_{\rm B}\hat{{\cal R}} ={\rm tr}_{{\rm S},{\rm B}}{\hat{{\cal D}}}{,\qquad}\hat{R}_{{\rm B}
}={\rm {\rm tr}}_{{\rm S},\,{\rm M}}{\hat{{\cal D}}
}{\rm ,\qquad}\hat{D}={\rm {\rm tr}}_{{\rm B}}{\hat
{{\cal D}}}{\rm  .} \label{reducedDM=}
\EEA 
The expectation value of any observable $\hat A$ pertaining, for instance, to the subsystem S + M of S + A 
(including products of spin operators $\hat s_a$ and $\hat\sigma^{(n)}_a$ ) can equivalently be evaluated as 
$\langle\hat A\rangle={\rm tr}_{S + A} \hat{\cal D}\hat A$ or as $\langle\hat A\rangle={\rm tr}_{S + M} \hat{D}\hat A$.

As indicated above, the apparatus \textrm{A} is a large system,
treated by methods of statistical mechanics, while we need to follow in detail
the microscopic degrees of freedom of the system \textrm{S} and their
correlations with \textrm{A}. To this aim, we shall analyze the full state
${\hat{{\cal D}}}$ of the system into several sectors, characterized by the
eigenvalues of $\hat{s}_{z}$. Namely, in the two-dimensional eigenbasis of
$\hat{s}_{z}$ for ${\rm S}$, $\left\vert {\uparrow}\right\rangle $,
$\left\vert {\downarrow}\right\rangle $, with eigenvalues $s_{i}=+1$ for
$i=\uparrow$ and $s_{i}=-1$ for $i=\downarrow$, ${\hat{{\cal D}}}$ can be
decomposed into the four blocks

\begin{equation}
\hat{\cal D}=
\left( \begin{array}{ccc@{\ }r}
{\hat{\cal R}}_{\uparrow\uparrow}& {\hat{\cal R}}_{\uparrow\downarrow}\\
{\hat{\cal R}}_{\downarrow\uparrow}& {\hat{\cal R}}_{\downarrow\downarrow}
\end{array} \right)
{ ,} \label{CDelemts}
\end{equation}
where each ${\hat {\cal R}}_{ij}$ is an operator in the space of the apparatus.
We shall also use the partial traces (see again Fig. 3.2)

\begin{equation}
\mytext{\textcurrency Rij\textcurrency \qquad}
\hat{R}_{ij}={\rm {\rm tr}}_{{\rm B}}{\hat {\cal R}}_{ij},{\qquad
\qquad}\hat{D}={\rm {\rm tr}}_{{\rm B}}{\hat {\cal D}}=
\left( \begin{array}{ccc@{\ }r}
\hat{R}_{\uparrow\uparrow} & \hat{R}_{\uparrow\downarrow}\\
\hat{R}_{\downarrow\uparrow} & \hat{R}_{\downarrow\downarrow}
\end{array} \right)
\label{Rij}
\end{equation}
over the bath; each $\hat{R}_{ij}$ is an operator in the $2^{N}$-dimensional
space of the magnet. Indeed, we are not interested in the evolution of the
bath variables, and we shall eliminate ${\rm B}$ by relying on the weakness
of its coupling (\ref{ham4}) with ${\rm M}$, expressed by the dimensionless variable $\gamma\ll1$. The operators $\hat{R}_{ij}$
code our full statistical information about ${\rm S}$ and ${\rm M}$. 
We shall use the notation $\hat R_{ij}$ whenever we refer to S + M and $\hat R_{\rm M}$ when referring 
to M alone. Tracing also over M, we are, according to (\ref{reducedDM=}), left with

\begin{equation}
\hat r
=\left( \begin{array}{ccc@{\ }r}
{{r}}_{\uparrow\uparrow}& {{ r}}_{\uparrow\downarrow}\\
{{r}}_{\downarrow\uparrow}& {{r}}_{\downarrow\downarrow}
\end{array} \right)= 
r_{\up\up}\,|\hspace{-1mm}\uparrow\rangle\langle\uparrow\hspace{-1mm}| +
r_{\up\down}\,|\hspace{-1mm}\uparrow\rangle\langle\downarrow\hspace{-1mm}| +
r_{\down\up}\,|\hspace{-1mm}\downarrow\rangle\langle\uparrow\hspace{-1mm}| +
r_{\down\down}\,|\hspace{-1mm}\downarrow\rangle\langle\downarrow\hspace{-1mm}| .
 \label{relements}
\end{equation}
 The magnet ${\rm M}$ is thus described by $\hat{R}_{\rm M}=\hat
{R}_{\uparrow\uparrow}+\hat{R}_{\downarrow\downarrow}$, the system
${\rm S}$ alone by the matrix elements of $\hat{r}$, viz.  $r_{ij}={\rm tr}
_{{\rm M}}\hat{R}_{ij}$. The correlations of $\hat{s}_{z}$,
$\hat{s}_{x} $ or $\hat{s}_{y}$ with any function of the observables
$\hat{\sigma}_{a}^{\left(  n\right)  }$ ($a=x,y,z$ , $n=1$ , \ldots$N$) are
represented by $\hat{R}_{\uparrow\uparrow}-\hat{R}_{\downarrow\downarrow}$,
$\hat{R}_{\uparrow\downarrow}+\hat{R}_{\downarrow\uparrow}$, $i\hat
{R}_{\uparrow\downarrow}-i\hat{R}_{\downarrow\uparrow}$, respectively. The
operators $\hat{R}_{\uparrow\uparrow}$ and $\hat{R}_{\downarrow\downarrow}$
are hermitean positive, but not normalized, whereas $\hat{R}_{\downarrow
\uparrow}=\hat{R}_{\uparrow\downarrow}^{\dagger}$.
Notice that we now have from (\ref{reducedDM=}) --  (\ref{Rij})

\BEA 
\mytext{\textcurrency reducedDM2=\textcurrency \qquad}
&& r_{ij} ={\rm {\rm tr}}_{{\rm A}}{\hat{{\cal R}}_{ij}} ={\rm {\rm tr}}_{{\rm M}}{\hat{{R}}_{ij}}   {,\qquad}
\hat{{\cal R}} = {\hat {\cal R}}_{\uparrow\uparrow} +{\hat {\cal R}}_{\downarrow\downarrow},\qquad
\nn \\ &&
\hat{{R}}_{{\rm M}}={\hat{R}}_{\uparrow\uparrow} +{\hat{R}}_{\downarrow\downarrow}
{,\qquad}\hat{R}_{{\rm B}}= {\rm {\rm tr}}_{{\rm M}}({\hat {\cal R}}_{\uparrow\uparrow} +{\hat {\cal R}}_{\downarrow\downarrow}){\rm  .} 
\label{reducedDM2=}
\EEA 

All these elements are functions of the time $t$ which elapses from the beginning of the measurement at  $t=0$ when $\hat H_{\rm SA}$ 
is switched on to the final value  $t_{\rm f}$ that we will evaluate in section 7. 

To introduce further notation, we mention that the combined system S + A = S + M + B should 
for all practical purposes end up in
\footnote{The terms $|\hspace{-1mm}\uparrow\rangle\langle\downarrow\hspace{-1mm}|$${\cal R}_{\up\down}(t)$ 
and $|\hspace{-1mm}\downarrow\rangle\langle\uparrow\hspace{-1mm}|$${\cal R}_{\down\up}(t)$ 
are not strictly zero, in fact the trace of their product
 ${\cal R}_{\up\down}^\dagger(t){\cal R}_{\up\down}(t)$ 
is even conserved in time. But when taking traces to obtain physical observables, the wildly oscillating phase factors which they carry
 prevent any meaningful contribution. There is clearly a discrepancy between {\it vanishing mathematically} and {\it being irrelevant physically.}
 \label{footFAPP}}

\begin{equation}
\hat{\cal D}(t_{\rm f})=
\left( \begin{array}{ccc@{\ }r}
p_{\uparrow}{\hat{\cal R}}_{\Uparrow}& 0 \\
0  & p_{\downarrow}{\hat{\cal R}}_{\Downarrow}
\end{array} \right)  
=p_{\uparrow}\,|\hspace{-1mm}\uparrow\rangle\langle\uparrow\hspace{-1mm}|\otimes {\hat{\cal R}}_{\Uparrow} 
+ p_\downarrow\, |\hspace{-1mm}\downarrow\rangle\langle\downarrow\hspace{-1mm}|\otimes{\hat{\cal R}}_{\Downarrow}
= \sum_{i=\up,\down} p_i \,\scriptD_i ,
\label{CDelemtsfin}
\end{equation}
where ${\hat{\cal R}}_{\Uparrow}$ (${\hat{\cal R}}_{\Downarrow}$) is density matrix of the thermodynamically 
stable state of the magnet and bath, after the measurement, in which the magnetization is up, taking the value  $m_{\Uparrow}(g)$ 
 (down, taking the value $m_{\Downarrow}(g)$); these events should occur with probabilities $p_{\uparrow}$ and $p_{\downarrow}$, 
respectively\footnote{Notice that in the final state we denote properties of the tested system by $\uparrow,\downarrow$
and of the apparatus by $\Uparrow,\Downarrow$. In sums like (\ref{CDelemtsfin}) we will also use $i=\uparrow,\downarrow$, or sometimes $i=\pm1$.
\label{fng}}.
When, at the end of the measurement, the coupling $g$ is turned off ($g\to0$), the macroscopic magnet will relax to the nearby state having
$m_{\Uparrow}(0)\approx m_{\Uparrow}(g)$  (viz. $m_{\Downarrow}(0)\approx m_{\Downarrow}(g)$).  
The Born rule then predicts that $p_{\uparrow}={\rm tr}_{\rm S}\hat r(0)\Pi_{\uparrow}=r_{\uparrow\uparrow}(0)$ and 
$p_{\downarrow}=r_{\downarrow\downarrow}(0)$.

Since no physically relevant off-diagonal terms occur in (\ref{CDelemtsfin}), a point that we wish to explain, 
and since we expect B to remain nearly in its initial equilibrium state, 
we may trace out the bath, as is standard in classical and quantum thermal physics, without losing significant information.
It will therefore be sufficient for our purpose to show that the final state is\footnote{Being the trace of (\ref{CDelemtsfin}) over the bath, its off-diagonal terms vanish, 
see footnote \ref{footFAPP}.}

\begin{equation}
\hat{D}(t_{\rm f})=
\left( \begin{array}{ccc@{\ }r}
p_{\uparrow}{\hat{ R}}_{{\rm M}\Uparrow}& 0 \\
0  & p_{\downarrow}{\hat{ R}}_{{\rm M}\Downarrow}
\end{array} \right)  
=p_{\uparrow}\, |\hspace{-1mm}\uparrow\rangle\langle\uparrow\hspace{-1mm}|\otimes {\hat{R}}_{{\rm M}\Uparrow}+
 p_\downarrow\, |\hspace{-1mm}\downarrow\rangle\langle\downarrow\hspace{-1mm}|\otimes {\hat{ R}}_{{\rm M}\Downarrow},
\label{Delemtsfin}
\end{equation}
now referring to the magnet M and tested spin S alone.

Returning to Eq. (\ref{reducedDM2=}), we note that from any density operator $\hat{R}_{}$ of the magnet we 
can derive the {\it probabilities $P_{\rm M}^{\rm dis}\left(  m\right)$ for $\hat{m}$ to take the eigenvalues}
$m$, where ``dis'' denotes their discreteness. These $N+1$ eigenvalues,

\begin{equation}
\mytext{\textcurrency eig\textcurrency \qquad}
m=-1{\rm ,\qquad}-1+\frac{2}
{N}{\rm ,\qquad}\ldots{\rm,\qquad}1-\frac{2}{N}{\rm ,\qquad}1{\rm ,}
\label{eig}
\end{equation}
have equal spacings $\delta m=2/N$ and multiplicities

\BEA
\mytext{\textcurrency deg\textcurrency \qquad}
\label{deg}
G\left(  m\right) =\frac{N!}{\left[  \frac{1}{2}N\left(  1+m\right)  \right]  !\left[  \frac{1}{2}N\left(  1-m\right)  \right]  !}
=e^{S(m)}
\EEA
The entropy reads for large $N$

\BEA \hspace{-1cm}
S(m)= 
N\left(-\frac{1+m}{2} \ln \frac{1+m}{2} -\frac{1-m}{2} \ln \frac{1-m}{2}\right)+\log \sqrt{\frac{2}{\pi N\left(  1-m^{2}\right)  }}
\EEA
Denoting by $\delta_{\hat{m},m}$ the projection operator on the subspace $m$
of $\hat{m}$, the dimension of which is $G\left(  m\right)  $, we have

\BEQ
P_{\rm M}^{\rm dis}\left(  m,t\right)  ={\rm tr}_{\rm M}\hat{R}_{\rm M}(t)\delta_{\hat{m},m}.
\label{R->P}
\EEQ 
where the superscript ``dis'' denotes that $m$ is viewed as a {\it discrete} variable, over which sums can be carried out.
In the limit $N\gg1$,  where $m$ becomes basically a {\it continuous} variable, we shall later work with the functions $P_{\rm M}(m,t)$, defined as

\begin{equation}  
P_{\rm M}(m,t)=\frac{N}{2}P_{\rm M}^{\rm dis}(m,t),\qquad \int_{-1}^1 \d m\,P_{\rm M}(m,t)=\sum_m P_{\rm M}^{\rm dis}(m,t)=1,
\label{Pdis2cont}
\end{equation}
that have a finite and smooth limit for $N\to\infty$.
A similar relation will hold between $P_{\up\up}^{\rm dis}(m,t)$ and $P_{\up\up}(m,t)$, to be encountered further on.

\subsubsection{Initial state}

\label{section.3.3.3}
 
 In order to describe an unbiased measurement, 
${\rm S}$ and ${\rm A}$ are statistically independent in the initial state  , which is expressed by
${\hat {\cal D}}\left(  0\right)  =\hat{r}\left( 0\right)  \otimes{\hat {\cal R}}\left(  0\right) $.
The $2\times2$ density matrix $\hat{r}\left(  0\right)  $ of \textrm{S} is arbitrary; 
 by the measurement we wish to gain information about it.
It has the form (\ref{relements}) with elements
$r_{\uparrow\uparrow}\left(  0\right)  $, $r_{\uparrow\downarrow}\left(
0\right)  $, $r_{\downarrow\uparrow}\left(  0\right)  $ and $r_{\downarrow
\downarrow}\left(  0\right)  $ satisfying the positivity and hermiticity conditions

\BEA
&& r_{\uparrow\uparrow}\left( 0\right)  +r_{\downarrow\downarrow}\left(  0\right)  =1,
\qquad
r_{\uparrow\downarrow}\left(  0\right)  =r_{\downarrow\uparrow}^{\ast}\left(0\right)  {\rm ,\qquad}
\nn\\&&
r_{\uparrow\uparrow}\left(  0\right)  r_{\downarrow
\downarrow}\left(  0\right)  \geq r_{\uparrow\downarrow}\left(  0\right)
r_{\downarrow\uparrow}\left(  0\right)  {\rm  .} \label{rij0}
\EEA 

At the initial time, the bath is set into equilibrium at the temperature\footnote{We use units where Boltzmann's constant  $k_B$ is equal to one; 
otherwise, $T$ and $\beta=1/T$ should be replaced throughout by  $k_BT$ and $1/k_B T$, respectively.} $T=1/\beta$.
The corresponding density operator is, 
\begin{equation}
\mytext{\textcurrency RB0=\textcurrency \qquad}
\hat{R}_{{\rm B}}\left(0\right)  =\frac{1}{Z_{{\rm B}}}e^{-\beta\hat{H}_{{\rm B}}}{\rm  ,}
\label{RB0=}%
\end{equation}
where $\hat{H}_{{\rm B}}$ is given in Appendix A and $Z_{\rm B}$ is the partition function. 
The connection between the initial state of the bath and its spectrum (\ref{Ktilde}) is described in Appendix B.

According to the discussion in section~2.1.1, the initial density operator
$\hat{{\cal R}}\left(  0\right)  $ of the apparatus describes the magnetic
dot in a metastable paramagnetic state and a bath. As justified below, we take for it the
factorized form
\begin{equation}
\mytext{\textcurrency RA0=\textcurrency \qquad}
{\hat {\cal R}}\left(
0\right)  =\hat{R}_{\rm M}\left(  0\right)  \otimes\hat{R}_{{\rm B}
}\left(  0\right)  {\rm  ,} \label{RA0=}
\end{equation}
where the bath is in the Gibbsian equilibrium state (\ref{RB0=}), at the temperature
$T=1/\beta$ {\it lower} than the transition temperature of ${\rm M}$, while the
magnet with Hamiltonian (\ref{HA}) is in a paramagnetic equilibrium state at a temperature $T_{0}
=1/\beta_{0}$ {\it higher} than its transition temperature:
\begin{equation}
\mytext{\textcurrency HF0\textcurrency \qquad}
\hat{R}_{\rm M}\left(0\right)  =\frac{1}{Z_{{\rm M}}}e^{-\beta_{0}\hat{H}_{{\rm M}}}{\rm  .}
\label{HF0}
\end{equation}

How can the apparatus be actually initialized in the non-equilibrium state
(\ref{RA0=}) at the time $t=0$? This \textit{initialization} takes place
during the time interval $-\tau_{{\rm init}}<t<0$. The apparatus is first
set at earlier times into equilibrium at the temperature $T_{0}$. Due to the
smallness of $\gamma$, its density operator is then factorized and
proportional to $\exp[{-\beta_{0}(  \hat{H}_{{\rm M}}+\hat{H}_{{\rm B}})  }]$. At the time $-\tau_{{\rm init}}$ the phonon bath
is suddenly cooled down to $T$. We shall evaluate in \S~5 the
{\it relaxation time} of ${\rm M}$ towards its equilibrium ferromagnetic states
under the effect of ${\rm B}$ at the temperature $T$. 
Due to the weakness of the coupling $\gamma$, this time 
this time is long and {\it dominates the duration of the experiment.} 
We can safely assume $\tau_{{\rm init}}$ to be
much shorter than this relaxation time so that ${\rm M}$ remains unaffected
by the cooling. On the other hand, the quasi continuous nature of the spectrum
of ${\rm B}$ can allow the phonon-phonon interactions (which we have
disregarded when writing (\ref{Hbath})) to establish the equilibrium of
${\rm B}$ at the temperature $T$ within a time shorter than $\tau
_{{\rm init}}$. It is thus realistic to imagine an initial state of the
form (\ref{RA0=}).

An alternative method of initialization consists in applying to the magnetic dot a {\it strong radiofrequency field}, which acts on M but not on B. 
The bath can thus be thermalized at the required temperature, lower than the transition temperature of M, while the populations of spins of M 
oriented in either direction are equalized. The magnet is then in a paramagnetic state, as if it were thermalized at an {\it infinite} temperature 
$T_0$ in spite of the presence of a cold bath. In that case we have the initial state (see Eq. (\ref{sigmaxyz0}))

\BEQ\label{purePM}
 \hat R_{\rm M}(0)=\frac{1}{2^N}\prod_{n=1}^N\hat\sigma_0^{(n)}.
\EEQ

The initial density operator (\ref{HF0}) of ${\rm M}$ being simply a
function of the operator $\hat{m}$, we can characterize it as in (\ref{R->P})
by the probabilities $P_{\rm M}^{\rm dis}\left(  m,0\right)  $ for $\hat{m}$ to take the values
(\ref{eig}). Those probabilities are the normalized product of the
degeneracy (\ref{deg}) and the Boltzmann factor,

\BEA 
\label{bzf} 
P_{\rm M}^{\rm dis}(m,0)&=&\frac{1}{Z_0} G(m)\exp\left[\frac{NJ}{4T_0} m^4 \right], 
\quad Z_0=\sum_m  G(m)\exp\left[\frac{NJ}{4T_0}m^4\right].
\EEA 
For sufficiently large $N$, the distribution $P_{\rm M}\left( m,0\right) =\half NP_{\rm M}^{\rm dis}(m,0) $ is peaked around $m=0$, with the Gaussian shape

\begin{equation}
\mytext{\textcurrency Pm0\textcurrency \qquad}
P_{\rm M}\left(  m,0\right)  \simeq
\frac{1}{\sqrt{2\pi}\,\Delta m}e^{-m^{2}/2\Delta m^{2}}.
\label{Pm0}
\end{equation}
This peak, which has a narrow width of the form
\begin{equation}
\mytext{\textcurrency DELTAm\textcurrency \qquad}
\Delta m=\sqrt{\left\langle
m^{2}\right\rangle }=\frac{1}{\sqrt{N}}{\rm  ,} \label{DELTAm}
\end{equation}
involves a large number, of order $\sqrt{N}$, of eigenvalues (\ref{eig}), so
that the spectrum can  be treated as a continuum (except in section 6.3).

 \subsubsection{Ferromagnetic equilibrium states of the magnet}

\label{section.3.3.4}
 
The measurement will drive M from its initial metastable state to one of its stable ferromagnetic states.
The final state (\ref{CDelemtsfin}) of ${\rm S}+{\rm A}$ after measurement will thus 
involve the two ferromagnetic equilibrium states $\hat{\cal R}_{i}$, $i=\,\Uparrow$ or $\Downarrow$. 
As above these states $\hat {\cal R}_{i}$ of the apparatus factorize, in the weak coupling limit
($\gamma \ll1$), into the product of (\ref{RB0=}) for the bath and a  ferromagnetic equilibrium state $\hat{R}_{{\rm M}i}$ 
for the magnet M.  The point of this section is to study the properties of such equilibrium states, whose temperature $T=1/\beta$ is induced by the bath.

\begin{figure}[h!]
\label{figABN3}
\centerline{\includegraphics[width=8cm]{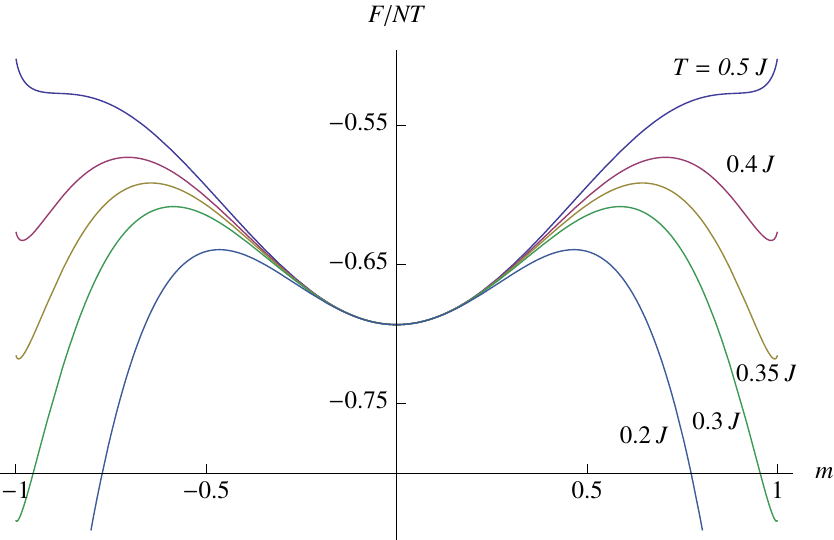}}
\caption{The free energy $F$ in units of $NT$, evaluated from Eq.  (\ref{F=}) with $h=0$, as function of the 
magnetization $m$ at various temperatures. There is always a local paramagnetic minimum at $m=0$. A first-order transition occurs at $T_{\rm c}=0.363J_4$, 
below which the ferromagnetic states associated with the minima at $\pm \,m_{\rm F}$ near $\pm\,1$
become the most stable.  }
\end{figure}

 Let us thus consider the equilibrium state of M, which depends on $\beta$ and on its Hamiltonian

\begin{equation}
\hat{H}_{{\rm M}}=-Nh\hat m-NJ\frac{\hat m^4}{4}{,} \label{HMh=}%
\end{equation}
where we introduced an external field $h$ acting on the spins of the apparatus for latter convenience.\footnote{In section 5 we shall identify 
$h$ with $+g$ in the sector $\hat R_{\up\up}$ of $\hat D$, or with $-g$ in its sector $\hat R_{\down\down}$, where $g$ is the coupling between S and A.}
As in (\ref{R->P}) we characterize the canonical equilibrium density operator of the magnet
$\hat{R}_{\rm M} =(1/Z_{\rm M})\exp[-\beta\hat H_{\rm M}]$, 
 which depends only on the operator $\hat{m}$, by the probability distribution


 \begin{equation}
P_{\rm M}\left(  m\right)  =
\frac{\sqrt{N}}{Z_{{\rm M}}\sqrt{8\pi }}e^{-\beta F\left(  m\right)  }{\rm  ,}
\label{Peq}%
\end{equation}
 where $m$ takes the discrete values $m_i$ given by (\ref{eig}); the exponent of  (\ref{Peq}) introduces the {\it free energy function}
                                    
\begin{equation}
F\left(  m\right)  =-NJ\frac{m^4}{4}-Nhm
+NT\left(\frac{1+m}{2} \ln \frac{1+m}{2} +\frac{1-m}{2} \ln \frac{1-m}{2}\right),
 \label{F=}
\end{equation}
which arises from the Hamiltonian (\ref{HMh=}) and from the multiplicity $G(m)$ given by (\ref{deg}). 
It is displayed in fig. \ref{figABN3}.
The distribution (\ref{Peq}) displays narrow peaks at the minima of $F\left(  m\right)  $, and
the \textit{equilibrium free energy} $-T\ln Z_{{\rm M}}$ is equal for large $N$ to the absolute minimum of (\ref{F=}). The function $F\left(  m\right)  $
reaches its extrema at values of $m$ given by the self-consistent equation

\begin{equation}
m=\tanh\left[  \beta\left(  h+J m^{3}\right)  \right]  {\rm  .} \label{MF}
\end{equation}
In the vicinity of a minimum of $F\left(  m\right)  $ at $m=m_{i}$, the probability
$P_{\rm M}\left(  m\right)  $ presents around each $m_{i}$ a nearly Gaussian peak, given within normalization by%

\begin{eqnarray}
P_{{\rm M}i}\left(  m\right)   \propto
\exp\left\{ -\frac{N}{2} \left[  \frac{1}{1-m_{i}^{2}}- 3\beta J m_i^2\right]  (  m-m_i)^2
\right\}.
  \label{Pim}%
\end{eqnarray}
This peak has a width of order $1/\sqrt{N}$ and a weak asymmetry. The possible values
of $m$ are dense within the peak, with equal spacing $\delta m=2/N$. With each
such peak $P_{{\rm M}i}\left(  m\right)  $ is associated through (\ref{R->P}), (\ref{Pdis2cont}), a
density operator $\hat{R}_{i}$ of the magnet ${\rm M}$ which may describe a locally stable equilibrium. 
Depending on the values of $J$  and on the temperature, there may exist one, two or three such locally stable states. 
We note the corresponding average magnetizations $m_i$, for arbitrary $h$,  as $m_{\rm P}$ for a paramagnetic state and as 
$m_\Uparrow$ and $m_\Downarrow$  for the ferromagnetic states, with $m_\Uparrow>0$, $m_\Downarrow<0$.
We also denote as $\pm m_{\rm F}$ the ferromagnetic magnetizations for $h=0$. 
When $h$ tends to $0$ (as happens at the end of the measurement where we set $g\to0$),  $m_{\rm P}$ tends to $0$, 
$m_\Uparrow$ to $+m_{\rm F}$ and $m_\Downarrow$ to $-m_{\rm F}$, namely

\BEA \label{mUpDown}
&& m_\Uparrow(h>0)>0,\qquad  m_\Downarrow(h>0)<0,\qquad 
m_\Uparrow(-h)=-m_\Downarrow(h),\qquad   \nn\\&& 
m_{\rm F}=m_\Uparrow(h\hspace{-0.5mm}\to\hspace{-0.5mm}+0)=-m_\Downarrow(h\hspace{-0.5mm}\to\hspace{-0.5mm}+0). 
\EEA

For $h=0$, the system ${\rm M}$ is invariant under change of sign of $m$.
This invariance is spontaneously broken below some temperature. 
The two additional ferromagnetic peaks   
 $P_{{\rm M}\Uparrow}\left(  m\right)  $ and $P_{{\rm M}\Downarrow}\left(  m\right)  $
appear around $m_{\Uparrow}=m_{{\rm F}}=0.889$ and $m_{\Downarrow
}=-m_{{\rm F}}$ when the temperature $T$ goes below $0.496J$. As $T$
decreases, $m_{{\rm F}}$ given by $m_{{\rm F}}=\tanh\beta J m_{{\rm F}%
}^{3}$ increases and the value of the minimum $F\left(  m_{{\rm F}}\right)
$ decreases; the weight (\ref{Peq}) is transferred from $P_{{\rm M}0}\left(m\right)$ to 
$P_{{\rm M}\Uparrow}\left(  m\right)  $ and $P_{{\rm M}\Downarrow}\left(m\right)  $. 
A first-order transition occurs when $F\left(  m_{{\rm F} }\right)  =F\left(  0\right)  $, for $T_{\rm c}=0.363J$ and $m_{{\rm F}}=0.9906$,
from the paramagnetic to the two ferromagnetic states, although the
paramagnetic state remains locally stable. The spontaneous magnetization
$m_{{\rm F}}$ is always very close to $1$, behaving as $1-m_{{\rm F}}\sim2\exp({-2J/T})$.

Strictly speaking, the canonical equilibrium state of ${\rm M}$ below the
transition temperature, characterized by (\ref{Peq}), has for $h=0$
and finite $N$ the form

\BEQ \label{RMeq}
\hat{R}_{{\rm Meq}}=\frac{1}{2}(  \hat{R}_{{\rm M }\Uparrow}+\hat{R}_{{\rm M}\Downarrow}\,).
\EEQ
However this state is not necessarily the one reached at the end of a relaxation process governed by the bath ${\rm B}$, when a field $h$, even weak, is present: this field acts
as a source which breaks the invariance. The determination of the state $\hat{R}_{\rm M}\left(  t_{{\rm f}}\right)  $ reached at the end of a
relaxation process involving the thermal bath ${\rm B}$ and a weak field $h$ {\it requires a dynamical study} which will be worked out in section 5. 
This is related to the ergodicity breaking: if a weak field is applied, then switched off, the full canonical state (\ref{RMeq})  is still recovered, but only after an unrealistically 
long time (for $N\gg 1$).  For finite times the equilibrium state of the magnet is to be found by restricting the full canonical state (\ref{RMeq})  to its component
having a magnetization with the definite sign determined by the weak external field. This is the essence of the spontaneous symmetry
breaking. However, for our situation this well-known recipe should be supported by dynamical considerations,
since we have to show that the thermodynamically expected states will be reached dynamically.

In our model of measurement, the situation is similar, though slightly more
complicated. The system-apparatus coupling (\ref{HSA}) plays the r\^{o}le
of an operator-valued source, with eigenvalues behaving as a field $h=g$ or
$h=-g$. We shall determine in section~6 towards which state ${\rm M}$ is
driven under the conjugate action of the bath ${\rm B}$ and of the system
${\rm S}$, depending on the parameters of the model.

{
\begin{figure}[h!]
\label{figABN4}
\centerline{\includegraphics[width=8cm]{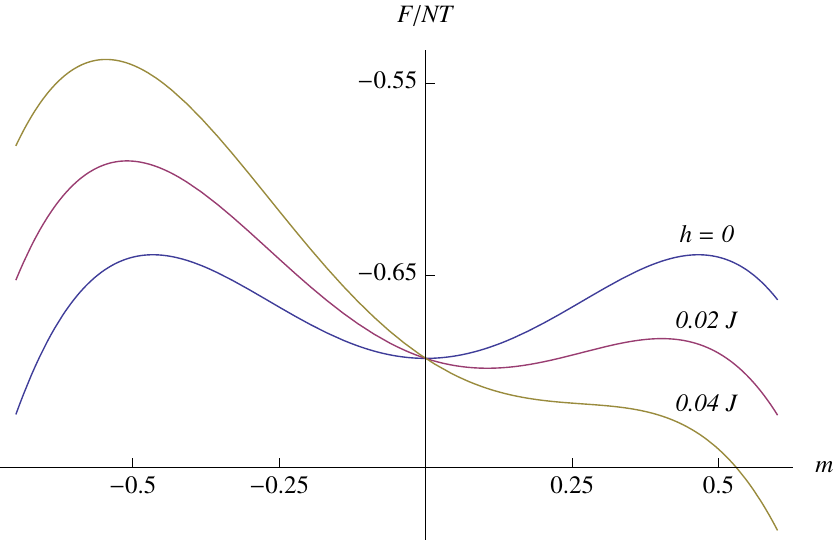}}
\caption{The effect of a positive field $h$ on $F(m)$ for $q=4$ at temperature $T=0.2J$.
As $h$ increases the paramagnetic minimum $m_{\rm P}$ shifts towards positive $m$. At the critical field $h_{\rm c}=0.0357J$ 
this local minimum disappears, and the curve has an inflexion point with vanishing slope at $m=m_{\rm c}=0.268$.  
For larger fields, like in the displayed case $g=0.04J$, the locally stable paramagnetic state disappears, and there remain only 
the two ferromagnetic states, the most stable one with positive magnetization $m_\Uparrow\simeq1$ and the metastable one
with negative magnetization $m_\Downarrow\simeq-1$.}
\end{figure}

As a preliminary step, let us examine here the effect on the free energy (\ref{F=}) of a small positive field $h$. Consider first the minima of
$F\left(  m\right)$  \cite{ll_stat,lavis}.
 The two ferromagnetic minima $m_{\Uparrow}$ and $m_{\Downarrow}$ given by (\ref{MF}) are slightly shifted away from
$m_{{\rm F}}$ and $-m_{{\rm F}}$, and $F\left(  m_{\Uparrow}\right) -F\left(  m_{{\rm F}}\right)  $ behaves as $-Nhm_{{\rm F}}$. Hence, as soon as 
$\exp\{{-\beta\left[  F\left(  m_{\Uparrow}\right)  -F\left(m_{\Downarrow}\right)  \right]  }\}\sim \exp({2\beta Nhm_{{\rm F}}})\gg1$, 
only the single peak $P_{{\rm M}\Uparrow}\left(  m\right)  $ around $m_{\Uparrow}\simeq
m_{{\rm F}}$ contributes to (\ref{Peq}), so that the canonical equilibrium
state of ${\rm M}$ has the form $\hat{R}_{{\rm Meq}}=\hat{R}_{{\rm M}\Uparrow}$. The shape of $F\left(  m\right)  $ will
also be relevant for the dynamics. 
If $h$ is sufficiently small, $F\left( m\right)  $ retains its paramagnetic minimum, the position of which is shifted as $m_{\rm P}\sim h/T$; the paramagnetic
state $\hat{R}_{\rm M}\left(  0\right)  $ remains locally stable. It may decay towards a stable ferromagnetic state only
through mechanisms of thermal activation or quantum tunneling, processes with very large characteristic times, of exponential
order in $N$. 
{In such cases A is not a good measuring apparatus}.
 However, there is a threshold $h_{{\rm c}}$ above which this paramagnetic minimum of $F\left(  m\right)  $, which
then lies at $m=m_{{\rm c}}$, disappears. The value of $h_{{\rm c}}$ is found by eliminating $m=m_{{\rm c}}$
between the equations ${\rm d}^{2}F/{\rm d}m^{2}=0$ and ${\rm d} F/{\rm d}m=0$. 
We find  $2m_{{\rm c}}^{2}=1-\sqrt{1-4T/3J}$, $h_{{\rm c}}=\frac{1}{2}T\ln[(1+m_{\rm c})/(1-m_{\rm c})]-Jm_{{\rm c}}^{3}$. 
At the transition temperature $T_{\rm c}=0.363J$,  we have $m_{{\rm c}}=0.375$ and $h_{{\rm c}}=0.0904J$;  for $T=0.2J$, we obtain
$m_{{\rm c}}=0.268$ and $h_{{\rm c}}=0.036J$;  for $T\ll J$, $m_{{\rm c}}$\ behaves as $\sqrt{T/3J}$ and $h_{{\rm c}}$\ as
$\sqrt{4T^{3}/27J}$. Provided $h>h_{{\rm c}}$, $F\left(  m\right)  $ has now a negative slope in the whole interval $0<m<m_{{\rm F}}$. 
We can thus expect, in our measurement problem, that the registration will take place in a reasonable delay for a first order transition if the coupling
 $g$ is larger than $h_{{\rm c}}$.\footnote{The set of conditions on parameters of A for being a good apparatus is reminiscent 
 of the requirements that realistic apparatuses have to fulfil. }

We have stressed already that the apparatus A should lie initially in a metastable state \cite{ll_stat,lavis}, and finally 
in either one of several possible stable states (see section 2 for other models of this type).  This suggests to take for A,
a quantum system that may undergo a phase transition with {\it broken invariance}. 
The initial state $\hat{{\cal R}}\left( 0\right) $ of ${\rm A}$\ is the metastable phase with unbroken invariance. The states $\hat{{\cal R}}_{i}$ 
represent the stable phases with broken invariance, in each of which registration can be permanent. The symmetry between the outcomes
prevents any bias. 

The initial state ${\hat {\cal R}}\left(  0\right)  $\ of ${\rm A}$\ is the metastable paramagnetic state.
We expect the final state (\ref{Delemtsfin}) of S + A to involve for ${\rm A}$ the two stable ferromagnetic states
${\hat {\cal R}}_{i}$, $i=\,\uparrow$ or $\downarrow$, that we denote as $\hat {\cal R}_{\Uparrow}$ or $\hat {\cal R}_{\Downarrow}$, 
respectively.$^{\ref{fng}}$ 
The equilibrium temperature $T$\ will be imposed to ${\rm M}$\ by the phonon bath \cite{petr,Weiss} through
weak coupling between the magnetic and non-magnetic degrees of freedom. Within
small fluctuations, the order parameter (\ref{hatm=}) vanishes
in ${\hat {\cal R}}\left(  0\right)  $\ and takes two opposite values in
the states ${\hat {\cal R}}_{\Uparrow}$ and ${\hat {\cal R}}_{\Downarrow}
$, $A_{i}\equiv\left\langle \hat{m}\right\rangle _{i}$ equal to
$+m_{{\rm F}}$ for $i=\uparrow$ and to $-m_{{\rm F}}$ for $i=\downarrow$\footnote{Note that the values $A_{i}=\pm m_{{\rm F}}$, which we wish to come out
associated with the eigenvalues $s_{i}=\pm1$, are determined from equilibrium
statistical mechanics; they are not the eigenvalues of $\hat{A}\equiv\hat{m}$,
which range from $-1$ to $+1$ with spacing $2/N$, but thermodynamic expectation values around which small fluctuations of order $1/\sqrt{N}$ occur.
For low $T$  they would be close to $\pm 1$.}.
As in real magnetic registration devices, information will be stored by ${\rm A}$\ in the form
of the sign of the magnetization.

\renewcommand{\thesection}{\arabic{section}}
\section{Dynamical equations}
\setcounter{equation}{0}\setcounter{figure}{0}
\renewcommand{\thesection}{\arabic{section}.}
\label{section.3}

In this section we present the basic steps that lead us to solvable evolution equations.
The Hamiltonian $\hat{H}_{0}$ in the space ${\rm S}+{\rm M}$
gives rise to two Hamiltonians $\hat{H}_{\uparrow}$ and
$\hat{H}_{\downarrow}$ in the space ${\rm M}$,\ which according
to (\ref{HSA}) and (\ref{HM=}) are simply two functions of the observable $\hat{m}$, given by%

\begin{equation}
\hat{H}_{i}=H_{i}\left(  \hat
{m}\right)  =-gNs_{i}\hat{m}-N\frac{J}{4}\hat{m}^{4}{\rm  ,} \qquad (i=\up,\down)
 \label{Hi}
\end{equation}
with $s_{i}=+1$ (or $-1$) for $i=\,\,\uparrow$ (or $\downarrow$).
These Hamiltonians $\hat{H}_{i}$, which describe interacting spins
$\mathbf{\hat {\sigma}}^{\left(  n\right)  }$ in an external field
$gs_{i}$, occur in
(\ref{dD}) both directly and through the operators

\begin{equation}
\hat{\sigma}_{a}^{\left(
n\right)  }\left(  u,i\right)  =e^{-i\hat{H}_{i}u/\hbar}\hat{\sigma}
_{a}^{\left(  n\right)  }e^{i\hat{H}_{i}u/\hbar}{\rm  .} \label{sigmaui}
\end{equation}

The equation (\ref{dD}) for $\hat D(t)$ which governs the joint dynamics of ${\rm S}+{\rm M}$ 
thus reduces to the four differential equations in the Hilbert space of ${\rm M}$
(we recall that $i,j=\uparrow, \downarrow$ or $\pm1$):

\BEA
\label{dRij}
&&\frac{{\rm d}\hat{R}_{ij}(t)}{{\rm d}t}-\frac{ \hat{H}_{i}\hat{R}_{ij}(t)-\hat{R}_{ij}(t)\hat{H}_{j}}{i\hbar}  =  \\
&&\frac{\gamma}{\hbar^{2}}\int_{0}^{t}{\rm d}u\sum_{n,a}\left\{  K\left(  u\right)  \left[  \hat{\sigma}_{a}^{\left(n\right)  }
\left(  u,i\right)  \hat{R}_{ij}(t),\hat{\sigma}_{a}^{\left(n\right)  }\right]  +K\left(  -u\right)  
 \left[  \hat{\sigma}_{a}^{\left(n\right)  },\hat{R}_{ij}(t)\hat{\sigma}_{a}^{\left(  n\right)  }\left(u,j\right)  \right]  \right\}.  \nn
\EEA
The action of the bath is compressed in $K(u)$, which is related to its spectrum (defined in (\ref{Ktilde})) through a Fourier transform:
\begin{equation}
K(t)=\frac{1}{2\pi}\int_{-\infty}^{+\infty}{\rm d}\omega \,e^{i\omega t}\tilde K\left(  \omega\right),
\qquad \tilde{K}\left(  \omega\right)=\int_{-\infty}^{+\infty}{\rm d}t\ e^{-i\omega t}K\left(  t\right).
\label{TF}
\end{equation}

To obtain the left hand side from the Liouville-von Neumann equation (\ref{dD}) is a straightforward exercise, 
but the right hand side, giving the action of the bath to lowest order in $\gamma$,  
involves several subtle steps explained in ref.1, that we reproduce here in Appendix C.

\subsection{The Born rule}

Taking the trace of (\ref{dRij}) in the $2^N\times 2^N$ dimensional Hilbert space of the magnet 
and over the bath, 
and using that the trace over the commutators vanishes, one obtains

\BEQ
\label{fal}
{i\hbar} \frac{{\rm d}\hat{r}_{ij}(t)}{{\rm d}t}={\rm tr} \, (\hat{H}_{i}-\hat{H}_{j})\hat{R}_{ij}(t)
=-gN(s_i-s_j){\rm tr}\,\hat m \hat{R}_{ij}(t).
\EEQ
Thus for $i=j$ one gets the conservation $r_{\uparrow\uparrow}(t)=r_{\uparrow\uparrow}(0)$ and
$r_{\downarrow\downarrow}(t)=r_{\downarrow\downarrow}(0)$.
This is the {\it Born rule} stating that the probabilities for outcomes is given by the state at the beginning of the measurement.
It is exactly obeyed, so in this aspect  the Curie-Weiss model describes an ideal measurement.  
Various other features desired  for ideal measurements will be satisfied in good approximation under suitable conditions on the system parameters.

An equivalent but simpler way to derive the Born rule is to notice that $i\hbar \d \hat s_z/\d t=[\hat s_z,\hat H]=0$, so that 
$\hat s_z$ is conserved, and with it the diagonal part $\half(1+\langle\hat s_z\rangle\,\hat s_z)$ of the density matrix $\hat r(t)$ .

The off-diagonal terms $\hat{r}_{ij}$ with $i\neq j$, that is to say, $r_{\uparrow\downarrow}(t)$ and $r_{\downarrow\uparrow}(t)$ or, equivalently, $\langle s_x\rangle$ 
and $\langle s_y\rangle$,
do evolve and actually go to zero, as discussed next. 
In popular terms this is called ``disappearance of Schr\"odinger cat terms''. 
Eq. (\ref{fal}) shows that the principle culprit is  the coupling $g$ between tested spin and magnet,
not the ferromagnetic interaction nor the bath.
Hence this step is a dephasing, not a decoherence.

\renewcommand{\thesection}{\arabic{section}}
\section{Decay of off-diagonal terms} 
\setcounter{equation}{0}\setcounter{figure}{0}
\renewcommand{\thesection}{\arabic{section}.}
\label{section.3}

Focusing on the Curie-Weiss model, we present here a derivation of the processes
which first lead to truncation of the off-diagonal elements of the density operator and which prevent recurrences from occurring. 
We show in section 6 and Appendix D of ref. \citen{ABNOpus} that the interactions with strength $\sim J$ between the spins $\hat\sigma^{(n)}$ of M play little role here, so that we neglect them. 
We further assume that M lies initially in the most disordered state (\ref{purePM}), that we write out, using the notation (\ref{sigmaxyz0}),  as

\BEQ 
\hat{R}_{\rm M}(0)=\frac{1}{2^N}\hat\sigma_0^{(1)}\otimes\hat\sigma_0^{(2)}\otimes\cdots\otimes\hat\sigma_0^{(N)}.
\EEQ
Then, since the Hamiltonian $\hat{H}_{{\rm SA}}+\hat{H}_{{\rm B}}
+\hat{H}_{{\rm MB}}$ is a sum of independent contributions associated with
each spin  {\boldmath{$\hat\sigma$}}$^{(n)}$, it can be shown from the Liouville-von Neumann equation (\ref{dD}) that, 
due to neglect of the coupling $J$,
the spins of M behave independently at all times, and that  the off-diagonal block $\hat{R}_{\uparrow\downarrow}(t)$ of
$\hat{D}(t)$ has the form
\begin{equation}
\hat{R}_{\uparrow\downarrow}(t)=r_{\uparrow\downarrow}(0)\,\hat{\rho}^{(1)}(t)\otimes\hat{\rho}
^{(2)}(t)\otimes\cdots\otimes\hat{\rho}^{(N)}(t){ ,} \label{Rfact}
\end{equation}
where $\hat{\rho}^{(n)}(t)$ is a $2\times2$ matrix in the Hilbert space of the spin 
 {\boldmath{$\hat\sigma$}}$^{(n)}$.
 This matrix will depend on $\hat{\sigma
}_{z}^{(n)}$ but not on $\hat{\sigma}_{x}^{(n)}$ and $\hat{\sigma}_{y}^{(n)}$,
and it will neither be hermitean nor normalized, except for $t=0$ where it equals $\half \hat\sigma_0^{(n)}$.

\subsubsection{Dephasing}

The first step in the dynamics of the off-diagonal terms happens at times where the bath is still inactive, 
the only active term in the Hamiltonian being the coupling to the tested spin.
Here spin $n$ processes as

\begin{equation}
\label{rhoneq0}
\frac{{\rm d}\hat{\rho}^{(n)}(t)}{{\rm d}t}=\frac{2ig}{\hbar}\hat{\rho}^{(n)}\hat{\sigma}_{z}^{(n)}
\end{equation}
with solution $\hat{\rho}^{(n)}(t)=\half \exp(2igt\hat\sigma_z^{(n)}/\hbar)=\half{\rm diag}[\exp({2igt/\hbar}),\exp({ -2igt/\hbar})]$.
One can easily deduce the related $P_{\uparrow\downarrow}(m)$ defined by (\ref{Rfact}) and  (\ref{R->P}).
Using that result or directly from  (\ref{Rfact}) it is simple to show that

\BEQ 
 r_{\uparrow\downarrow}\left(t\right)  =r_{\uparrow\downarrow}\left(  0\right)  \left( \cos\frac{2gt}{\hbar} \right)^N
\EEQ
For large $N$ this expression decays quickly in time, 

\begin{equation}
r_{\uparrow\downarrow}\left(t\right)  =r_{\uparrow\downarrow}\left(  0\right)  
e^{-\left(  t/\tau_{{\rm trunc}}\right)  ^{2}}{\rm  ,} \label{rofft}
\end{equation}
or equivalently
\begin{eqnarray}
\mytext{\textcurrency stransv\textcurrency \qquad}
\left\langle \hat{s}
_{a}\left(  t\right)  \right\rangle  &  =&\left\langle \hat{s}_{a}\left(
0\right)  \right\rangle e^{-\left(  t/\tau_{{\rm trunc}}\right)  ^{2}
}{, \qquad\qquad}(a=x{, }y){ ,}\label{stransv}\\
\label{slong}
\end{eqnarray}
where we introduced the truncation time 

\begin{equation}
\tau_{{\rm trunc}}\equiv
\frac{\hbar}{\sqrt{2} \ Ng\Delta m}=\frac{\hbar}{\sqrt{2N}\ \delta_{0}g}{ .}
\label{taured}
\end{equation}
Although $P_{\uparrow\downarrow}(m,t)$ is merely an oscillating function of $t$ for each value of $m$, the summation over $m$ has 
given rise to extinction. This property arises from the dephasing that exists between the oscillations for different values of $m$. 
There are undesired recurrences, however, when $2gt/\hbar=n\pi$, $n=1,2,\cdots$, which can be suppressed by a spread in the coupling $g$ (see below)
or by the action of the bath.

\subsubsection{Decoherence}

It is generally believed that Schr\"odinger cat terms (here:  $\hat{r}_{\uparrow\downarrow}$ and $\hat{r}_{\downarrow\uparrow}$)
disappear due to a coupling to a bath (environment).  
However, we stress that the basis in which the off-diagonal blocks of the density matrix of S+M disappear is not selected by the interaction 
with the environment (here with the bath B), but by the coupling between S and M. Moreover, for the present model, we have seen in the previous section 
that the main phenomenon which lets the off-diagonal blocks decay rapidly is dephasing.
Here we look at the subsequent role of decoherence, while still neglecting $J$. 
We  leave open the possibility for the coupling $g_n$ to be random,  whence
the coupling between S and A reads  $\hat H_{\rm SA}=-\hat s_z\sum_{n=1}^Ng_n\hat \sigma_z^{(n)}$ instead of (\ref{HSA}).
Each factor $\hat{\rho}^{(n)}(t)$, initially equal to $\frac{1}{2}\hat\sigma_0^{(n)}$, evolves according to the same equation, since in absence
of $J$, the Hamiltonian is a sum of single apparatus-spin terms. 
It can be found by inserting the product structure (\ref{Rfact}) into (\ref{dRij}), or by taking the latter for $N=1$.
Let us denote $2g_n/\hbar=\Omega_n$. 
In the limit $J\to0$ one can show that

\BEQ 
\hspace{-4mm} \label{sigmanau}
\hat\sigma_x^{(n)}(u,i)&=&\cos\Omega_nu \, \hat\sigma_x^{(n)}-s_i\sin\Omega_nu\,\hat\sigma_y^{(n)},
\nn\\
\hat\sigma_y^{(n)}(u,i)&=&\cos\Omega_nu \, \hat\sigma_y^{(n)}+s_i\sin\Omega_nu\,\hat\sigma_x^{(n)},
\EEQ
while of course $\hat\sigma_z^{(n)}(u,i)=\hat\sigma_z^{(n)}$ is conserved.
Each $\hat \rho^{(n)}$ is only a function of $\Omega_n$ and $t$,  viz. $\hat \rho^{(n)}(t)=\hat\rho(\Omega_n,t)$,
having the diagonal form $\hat \rho(t)=\half[\rho_0(t)\hat\sigma_0+i\rho_3(t)\hat\sigma_3]$.

The effect of the bath is relevant only at times $t \gg \tau_T=\hbar/2 \pi T$, where
$\hat \rho(\Omega,t)$ evolves according to

\BEQ
\frac{{\rm d}\hat{\rho}(t)}{{\rm d}t}
=i\Omega\hat{\rho}\hat{\sigma}_{z}+\frac{2i\gamma \hat\sigma_z}{\hbar^2}\int_0^t\d u[K(u)+K(-u)](\rho_0\sin\Omega u-\rho_3\cos\Omega u).
\EEQ
This encodes the scalar equations

\BEQ
\dot\rho_0=-\Omega\rho_3,\qquad 
\dot \rho_3=\Omega\rho_0+\mu\rho_0-2\lambda \rho_3
\EEQ
where 

\BEQ \label{lamtmut}
\lambda &=&\frac{2\gamma}{\hbar^2}\int_0^t\d u[K(u)+K(-u)]\cos\Omega u,\nn\\
\mu &=&\frac{4\gamma }{\hbar^2}\int_0^t\d u[K(u)+K(-u)]\sin\Omega u.
\EEQ
For times larger than $\tau_T=2\pi\hbar/T$ the integrals may be taken to infinity, so that $\lambda$ and $\mu$ become constants.
The Ansatz $\rho_0=A \exp xt$, $\rho_3=C\exp xt$ then yields

\BEQ
\label{Omp=}
x_\pm=-\lambda \pm i\Omega',\qquad \Omega'=\sqrt{\Omega^2+\Omega \mu-\lambda^2}.
\EEQ
and, taking into account the initial conditions, the solution reads

\BEQ
\rho_0=\frac{\Omega'\cos\Omega't+\lambda\sin\Omega't}{\Omega'}e^{-\lambda t},\qquad
\rho_3=\frac{\Omega+\mu}{\Omega'}\sin\Omega't\, e^{-\lambda t}.
\EEQ
For small $\gamma$ they imply 

\BEQ 
\label{outrho}
\hat\rho(t)=\half e^{-\lambda t+i\Omega\hat\sigma_z t}.
\EEQ
For $t\gg \tau_T$ the coefficient $\lambda$ is equal to

\BEQ
\lambda\equiv \lambda(\infty)=\frac{\gamma}{\hbar^2}[\tilde K(\Omega)+\tilde K(-\Omega)]
=\frac{\gamma\Omega}{4}\coth\half\beta\hbar\Omega=\frac{\gamma g_n}{2\hbar}\coth\frac{g_n}{T},
\EEQ
where we could neglect the cutoff $\Gamma$. 
The coefficient $\mu$, only occurring as a small frequency shift in (\ref{Omp=}),  is less simple.
After a few straightforward steps one has

\BEQ
\hspace{-4mm}
\mu(t)=\frac{\gamma}{2\pi}
\int_{0}^\infty \d \omega\,{\omega\,e^{-\omega/\Gamma}}{\coth\frac{\beta\hbar\omega}{2}}\left(
\frac{1-\cos(\omega-\Omega)t}{\omega-\Omega}- \frac{1-\cos(\omega+\Omega)t}{\omega+\Omega}\right).
\EEQ
Its $t\to\infty$ limit is obtained by dropping the cosines. Inserting $\coth=1+(\coth-1)$ and splitting the integral, one 
gets  from the first part $(\gamma\Omega/\pi)(\log\Gamma/\Omega-\gamma_E)$  with Euler's constant $\gamma_E=0.577215$,
while one may put $1/\Gamma\to0$ in the second part. 
In fact, a further splitting $\coth-1=(\tanh-1)+(\coth-\tanh)$ may be done to separate a possible logarithm in $\beta\hbar\Omega$, while
one may perform a contour integration in the last part.

By inserting (\ref{outrho}) into (\ref{Rfact}) and tracing out the pointer variables, one finds the transverse polarization of S as
\BEA
&&\half\langle\hat s_x(t)- i \hat s_y(t)\rangle\equiv {\rm tr}_{\rm S,A}\hat{\cal D}(t)\half(\hat s_x- i \hat s_y)=
r_{\uparrow\downarrow}(t)\equiv r_{\uparrow\downarrow}(0)\,{\rm Evol}(t),\qquad \EEA
where the temporal evolution is coded in 

\BEA\label{Evol}
\hspace{-6mm}
{\rm Evol}(t)\equiv
\left({ \prod}_{n=1}^N\cos\frac{2g_nt}{\hbar}\right)\,
\exp\left( - \sum_{n=1}^N\frac{\gamma g_n}{2 \hbar}\coth \frac{g_n}{T}\,t
\right).
 \EEA
To see what this describes, one can first take $g_n=g$, $\gamma=0$ and plot the factor $|{\rm Evol}(t)|$ 
from $t=0$ to $5\tau_{\rm recur}$, where $\tau_{\rm recur}= \pi\hbar/2g$ 
is the time after which $|r_{\uparrow\downarrow}(t)|$ has recurred to its initial value $|r_{\uparrow\downarrow}(0)|$. 
By increasing $N$, e.g.,  $N=1,2,10,100$,  one convince himself that the decay near $t=0$ becomes 
close to a Gaussian decay, over the characteristic time $\tau_\trunc=\hbar/\sqrt{2N}g$. One may demonstrate this analytically
by setting $\cos2g_nt/\hbar\approx\exp(-2g_n^2t^2/\hbar^2)$  for small $t$.
This time characterizes dephasing, that is, disappearance of the off-diagonal blocks of the density matrix while still phase coherent; 
we called it ``{\it truncation time}'' rather than ``{\it decoherence time}'' to distinguish it from usual decoherence, which is induced by a thermal environment
and coded in the second factor of Evol($t$).  

In order that the model describes a faithful quantum measurement, it is mandatory  that $|$Evol$|\ll1$ at $t=\tau_{\rm recur}$. 
 To this aim,  keeping $\gamma=0$, one can in the first factor of Evol decompose $g_n=g+\delta g_n$,  
where $\delta g_n$ is a small Gaussian random variable with $\langle\delta g_n\rangle=0$ and
 $\langle\delta g_n^2\rangle\equiv\delta g^2\ll g^2$, and average over the $\delta g_n$. 
The Gaussian decay (\ref{rofft}) will thereby be recovered, which already prevents recurrences.
One may also take e.g. $N=10$ or $100$, and plot the function to show this decay and to estimate the size of Evol at later times.

Next by taking $\gamma>0$ the effect of the bath in (\ref{Evol}) can be analyzed. For values $\gamma$ such that $\gamma N\gg1$ the bath
will lead to a suppression called {\it decoherence}, as is exemplified by the dependence on the bath temperature $T$.
It is ongoing, not once-and-for-all \cite{ABNOpus}.
 Several further  aspects can be easily considered now: Take all $g_n$ equal and plot the function Evol($t$); take a small spread in them and compare the results;
make the small-$g_n$ approximation $g_n \coth g_n/T\approx T$, and compare again. 

At least one of the two effects (spread in the couplings or suppression by the bath) should be strong enough to prevent 
recurrences, that is, to make $|r_{\uparrow\downarrow}(t)|\ll |r_{\uparrow\downarrow}(0)|$
at any time $t\gg \tau_\trunc$, including the recurrence times.\footnote{The condition {\it strong enough} poses constraints on the parameters
for the apparatus to function properly. In contrast, the interaction of the billions of solar neutrinos that pass
our body every second  is  {\it weak enough} to prevent the destruction of life.}
In the dynamical process for which each spin {\boldmath{$\hat\sigma$}}$^{(n)}$ of M independently
rotates and is damped by the bath, the truncation, which destroys the expectation values $\langle\hat s_a\rangle$ and all correlations 
$\langle\hat s_a\hat m^k(t)\rangle$ ($a=x$ or $y$, $k\ge 1$), arises from the precession of the tested spin $\hat s$
around the $z$-axis; this is caused by the conjugate effect of the many spins $\hat\sigma^{(n)}$ of M,  while the suppression of recurrences is either due
to dephasing if the $g_n$ are non-identical, or due to damping by the bath.

Finally, one may go back to the {\it time-dependent } expressions (\ref{lamtmut}) for $\lambda$ and $\mu$ 
and deduce how the initial growth at small $t$ can, for large $N$, already induce the decoherence \cite{ABNOpus}.

\renewcommand{\thesection}{\arabic{section}}
\section{Dynamics of the registration process}
\setcounter{equation}{0}\setcounter{figure}{0}
\renewcommand{\thesection}{\arabic{section}.}
\label{section.3}

\label{section.9.6.2}
\label{fin9.6.2}
The purpose of a measurement is the registration of the outcome, which can then be read off.
For the description of the registration process we need to study $P_{ii}(m,t)$ defined in terms of $\hat R_{ii}(t)$ in (\ref{R->P}).
The equations for $P_{ij}(m,t)$ follow from (\ref{dRij}) and are derived in Appendix B of ref. \citen{ABNOpus}.

The integrals over $u$ produce the functions   $\tilde{K}_{t>}\left(  \omega\right)$ and $  \tilde{K}_{t>}\left(  \omega\right)$

\begin{eqnarray}
\mytext{K>}\qquad 
  \tilde{K}_{t>}\left(  \omega\right)&  =&\int_{0}
^{t}{\rm d}ue^{-i\omega u}K\left(  u\right)  =\frac{1}{2\pi i}\int
_{-\infty}^{+\infty}{\rm d}\omega^{\prime}\tilde{K}\left(  \omega^{\prime}\right)
\frac{e^{i\left(  \omega^{\prime}-\omega\right)  t}-1}{\omega^{\prime}-\omega}
 {\rm  ,}
\label{K>}
\EEA 
and

\BEA
\mytext{K<}\qquad
{\rm   \tilde{K}_{t<}}\left(  \omega\right)  &=&\int_{0}
^{t}{\rm d}ue^{i\omega u}K\left(  -u\right)  =\int_{-t}
^{0}{\rm d}ue^{-i\omega u}K\left(  u\right) =\left[  \tilde{K}_{t>}\left(
\omega\right)  \right]  ^{\ast}
  {\rm  ,}\quad \label{K<}
\end{eqnarray}
where $\omega$ takes, depending on the considered term, the values
$\Omega^+_\uparrow$, $\Omega^-_\uparrow$,  $\Omega^+_\downarrow$, $\Omega^-_\downarrow$, 
given by

\BEQ \label{Omidef}
\hbar\Omega^\pm_i(m)=H_i(m\pm\delta m)-H_i(m),\qquad (i=\up,\down),
\EEQ
in terms of the Hamiltonians (\ref{Hi}) and of the level spacing $\delta m=2/N$. They satisfy the relations
$\Omega_i^\pm(m\mp\delta m)=-\Omega_i^\mp(m)$.
The quantities (\ref{Omidef}) are interpreted as excitation energies of the magnet M arising from the flip of one of its spins 
 in the presence of the tested spin S (with value $s_i$); the sign $+$ ($-$) refers to a down-up (up-down) spin flip. 
 Their explicit values are: 
 
\BEA
\label{ompmi=}
\hbar\Omega^\pm_i(m)=\mp2gs_i
+2J(\mp m^3-\frac{3m^2}{N}\mp\frac{4m}{N^2}-\frac{2}{N^3}),
\EEA
with $s_\uparrow =1$, $s_\downarrow =-1$. 

The operators $\hat\sigma^{(n)}_x$ and  $\hat\sigma^{(n)}_y$  which enter (\ref{dRij}) are shown in Appendix B to produce a flip of the spin 
{\boldmath{$\hat\sigma$}}$^{(n)}$, that is, a shift of the operator $\hat m$ into $\hat m\pm\delta m$. 
 We introduce the notations

\begin{eqnarray}
\Delta_{\pm}f\left(  m\right) =f\left(  m_{\pm}\right)  -f\left(  m\right),\qquad 
m_\pm=m\pm\delta m,\qquad \delta m=\frac{2}{N}  {\rm  .} \label{delta+-}
\end{eqnarray}

The resulting dynamical equations for $P_{ij}(m,t)$ take different forms for the diagonal and for
the off-diagonal components. On the one hand, the first \textit{diagonal
block} of $\hat{D}$ is parameterized by the \textit{joint probabilities}
$P_{\uparrow\uparrow}\left(  m,t\right)  $ to find ${\rm S}$ in $\left\vert
\uparrow\right\rangle $ and $\hat{m}$ equal to $m$ at the time $t $. 
In the Markov regime $t\sim  J/\gamma$ these
probabilities evolve according to

\BEA 
\label{dPdiag}
\frac{{\rm d}P_{\uparrow
\uparrow}\left(  m,t\right)  }{{\rm d}t}=\frac{\gamma N}{\hbar^{2}}&&\left\{ \,\,
\Delta_{+}\left[  \left(  1+m\right)  \tilde{K}\left(  \Omega_{\uparrow
}^{-}(m)\right)  P_{\uparrow\uparrow}\left(  m,t\right)  \right] \right.\\ &&\left. +\, \Delta
_{-}\left[  \left(  1-m\right)  \tilde{K}\left(  \Omega_{\uparrow}
^{+}(m)\right)  P_{\uparrow\uparrow}\left(  m,t\right)  \right]  \right\}  {\rm ,} 
\nn
\EEA 
with initial condition $P_{\uparrow\uparrow}\left(  m,0\right)
=r_{\uparrow\uparrow}\left(  0\right)  P_{\rm M}\left(  m,0\right)  $ given by
(\ref{Pm0}) and boundary condition $P_{\uparrow\uparrow}(-1-\delta m)=P_{\uparrow\uparrow}(1+\delta m)=0$;
 likewise for $P_{\downarrow\downarrow}\left(  m\right)$,  which involves the frequencies $\Omega^\mp_\down(m)$.
 The factor $\tilde K$ is introduced in Eq. (\ref{Ktilde}). On times $t\ll T/\gamma$, 
Eq. (\ref{dPdiag}) should actually involve the more complicated form  $\tilde{K}_{t}\left(  \omega\right)$, given by

\begin{equation}
\mytext{\textcurrency Ktomega\textcurrency \qquad}
\tilde{K}_{t}\left(\omega\right)
 \equiv\tilde{K}_{t>}\left(\omega\right)  +\tilde{K}_{t<}\left(  \omega\right) 
   =\int_{-t}^{+t}{\rm d}ue^{-i\omega u}K\left(  u\right)
=\int_{-\infty}^\infty \frac{{\rm d}\omega^{\prime}}{\pi}\frac{\sin\left(  \omega^{\prime
}-\omega\right)  t}{\omega^{\prime}-\omega}\tilde{K}\left(  \omega^{\prime
}\right)  {\rm  .} \label{Ktomega}
\end{equation}
This expression is real too and tends to $\tilde{K}\left(  \omega\right)$ 
at times $t$ larger than the range $\hbar/2\pi T$ of $K\left(  t\right) $ \cite{petr,Weiss} \footnote{Students with numerical skills may check
this by programming the integral; those with analytical skills may replace the cutoff factor  $\exp(-|\omega|/\Gamma)$ of $\tilde K(\omega)$ in (\ref{Ktilde})
by the quasi-Lorentzian $4\tilde\Gamma^4/(\omega^4+4\tilde\Gamma^4)$ and do a contour integral in the upper half plane. See also Appendix D of ref. \citen{ABNOpus}.},
as may be anticipated from the relation  $\sin[(\omega'-\omega)t]/(\omega'-\omega)\to \pi \delta(\omega'-\omega)$ for $t\to\infty$.
Fortunately, the dynamics of the relaxation process which moves the magnet from its initial paramagnetic phase to one of the stable ferromagnetic phases
takes place on times $t\sim T/\gamma$, after which $\tilde K_t(\omega)$  has relaxed to the simpler expression $\tilde K(\omega)$, 
so this evolution is to a very good approximation given by (\ref{dPdiag}). 
 This makes it possible to solve the difference equations (\ref{dPdiag}) numerically for $N=10$, $100$, $1000$ or larger.
 (As mentioned, it holds that $P=0$ for $m=1+2/N$ or $-1-2/N$.)
Figure  \ref{figABN11} presents the result at different times for $N=1000$.

\begin{figure}[h!]
\label{figABN11}
\includegraphics[width=8cm]{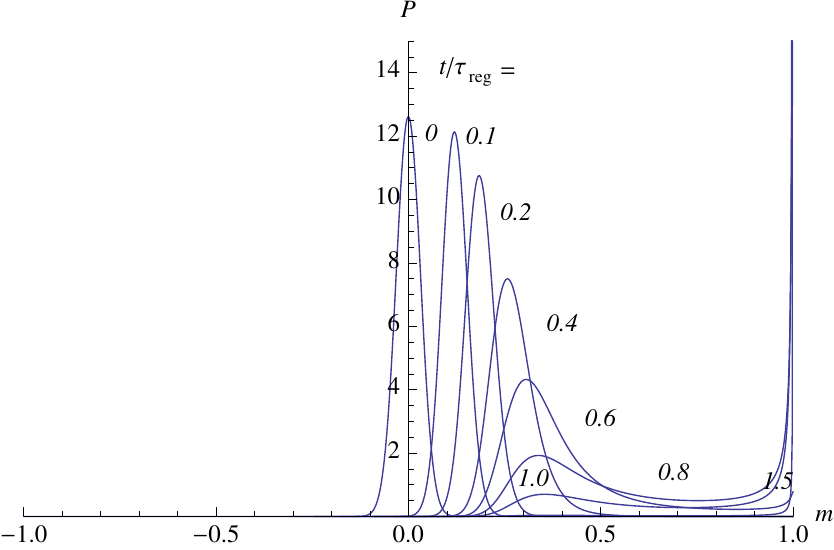}
\caption{
The registration process for quartic Ising interactions.
The probability density $P(m,t)=P_{\uparrow\uparrow}(m,t)/r_{\uparrow\uparrow}(0)$ 
as function of $m$ is represented at different times up to $t=1.5\,\tau_{\rm reg}$.
The parameters are chosen as $N=1000$,  $T=0.2J$ and $g=0.045J$ as in  Fig 7.4.
The time scale is here the registration time $\tau_{\rm reg}=38 \tau_J=38 \hbar/\gamma J$, which is large due to the existence of 
a bottleneck around $m_{\rm c}=0.268$. The coupling $g$ exceeds the critical value $h_{\rm c}=0.0357J$ needed for proper registration, 
but since $(g-h_{\rm c})/h_{\rm c}$ is small, the drift velocity has a low positive minimum at $0.270$ near $m_{\rm c}$ (Fig. 7.2). 
Around this minimum, reached at the time $\frac{1}{2}\tau_{\rm reg}$, the peak shifts slowly and widens much. 
Then, the motion fastens and the peak narrows rapidly, coming close to ferromagnetism around the time 
$\tau_{\rm reg}$, after which equilibrium is exponentially reached.}
\end{figure}

One may also proceed analytically.
It takes a few steps (see ref. \citen{ABNOpus})  to approximate (\ref{dPdiag}) for large $N$ by the Fokker-Planck equation

\begin{equation}
\mytext{\textcurrency dPdiag2\textcurrency \qquad}
\frac{\partial P_{\uparrow
\uparrow}}{\partial t}\approx\frac{\partial}{\partial m}\left[  -v\left(  m,t\right)  P_{\uparrow\uparrow}\right]  +\frac{1}{N}
\frac{\partial^2}{\partial m^2}\left[  w\left(  m,t\right) P_{\uparrow\uparrow}\right]  {\rm  ,}
\label{dPdiag2}
\end{equation}
where 

\begin{eqnarray}
\mytext{\textcurrency vupup\textcurrency \qquad}
v\left(m,t\right)   &  =& \frac{2\gamma}{\hbar^{2}}\left[  \left(  1-m
\right)  \tilde{K}_{t}\left(  -2\omega_{\uparrow}\right)  -\left(1+m
\right)  \tilde{K}_{t}\left(  2\omega_{\uparrow}\right)\right]
 {\rm  ,}\label{vupup}\\
\mytext{\textcurrency wupup\textcurrency \qquad}
w\left(m,t\right)   &  =& \frac{2\gamma}{\hbar^{2}}\left[  \left(  1-m\right)  \tilde
{K}_{t}\left(  -2\omega_{\uparrow}\right)  +\left(  1+m\right)  \tilde{K}
_{t}\left(  2\omega_{\uparrow}\right)  \right]  
{\rm  .} \label{wupup}
\end{eqnarray}

One would be inclined to leave out the diffusion term of order $1/N$.
Indeed, if we keep aside the shape  and the width of the probability distribution, which has a narrow peak 
for large $N$,  the center $\mu(t)$ of this peak moves according to the mean-field equation

\BEQ 
\label{mudot=}
\frac{\d\mu(t)}{\d t}=v[\mu(t)],
\EEQ
where $v(m)$ is the local drift velocity of the flow of $m$,

\BEQ
\label{v92}
v(m)=\frac{\gamma}{ \hbar} (g+Jm^{q-1}) \left(1 -m\,\coth \frac{g+Jm^{q-1}}{ T}\right).
\EEQ
This result can be derived by multiplying (\ref{dPdiag2}) by $m$ and integrating over it, while the narrowness of $P(m)$ 
around its peak at $\mu$ allows to replace $m$ by $\mu$ inside $v$.

If the coupling $g$ is large enough, the resulting dynamics will correctly describe the transition of the magnetization from 
the initial paramagnetic value $m=0$ to the final ferromagnetic value $m=m_{\rm F}$.
As a task, one can determine the minimum value of the coupling $g$ below  which the registration cannot take place.
Approaching this threshold from above, one observes the slowing down  of the process around the crossing of the bottleneck.

Focussing on $\mu(t)=\langle m(t)\rangle$
overlooks the broadening and subsequent narrowing of the profile at intermediate times, which is relevant for 
finite values of $N$. 
This can be studied by numerically solving  the time evolution of $P(m,t)$, i. e.,
the whole registration process, at finite $N$, taking in the rate equations  Eq. (\ref{dPdiag}) 
e.g. $N=10,\,100$ and $1000$. For the times of interest, $t\sim1/\gamma$, one is allowed to employ the simplified form of the
 rates that arise from setting $\tilde K_t(\omega)\to\tilde K(\omega)$ and employing (\ref{ompmi=}).  The relevant rate coefficients are 

\BEQ 
\label{Brate}
\frac{\gamma N}{\hbar^2}\tilde K(\omega)=
\frac{N\hbar\omega}{8J\,\tau_J}\left[\coth\left(\half\beta\hbar\omega \right)-1\right]\,
\exp\left(-\frac{|\omega|}{\Gamma}\right),
\EEQ
where the timescale $\tau_J=\hbar/\gamma J$ can be taken as a unit of time. The variable $\omega$ in $\tilde K(\omega)$ takes the values
 $\Omega^\pm _i$, with $i=j=\,\, \uparrow$ or $\downarrow$, which are explicitly given by (\ref{ompmi=}) in terms of the discrete variable $m$. 
 It can be verified that, for $\Gamma\gg J/\hbar$, the omission of the Debye cut-off in (\ref{Brate}) does not significantly affect the dynamics.

\renewcommand{\thesection}{\arabic{section}}
\section{The quantum measurement problem and the elements of its solution}
\setcounter{equation}{0}\setcounter{figure}{0}
\renewcommand{\thesection}{\arabic{section}.}
\label{section.3}

 In the measurement postulates of textbooks it is taken for granted that individual measurements yield individual outcomes.
However, on a theoretical level this is a non-trivial feature to be explained, know as the ``measurement problem''.

\subsection{Why the task is not achieved: the quantum ambiguity}

We have shown in Sections 4 and 5 that, for suitable values of the parameters entering the Hamiltonian, S+A ends up for the Curie--Weiss model in an 
equilibrium state represented by the density operator
	
\BEQ	\scriptD(\tf)=\sum_i p_i \hat r_i \otimes \scriptR_i  .   
\EEQ
The index $i$ takes two values associated with up or down spins; the weights $p_i$ are equal to the diagonal elements $r_{\up \up} (0)$ or
$r_{\down \down} (0)$ of the initial state of S; the states $\hat r_i$ of S are the projection operators $|\hspace{-1mm}\up\rangle\langle \up\hspace{-1mm}|$ or 
$|\hspace{-1mm}\down\rangle\langle \down\hspace{-1mm}|$ on the eigenspaces associated with the values +1 or -1 of $s_z$; the states $\scriptR_\Up$ or $\scriptR_\Down$ 
are  the ferromagnetic equilibrium states of A. This state (6.1) exhibits the required one-to-one correspondence between the eigenvalue of $\hat s$ and the indication
 of the pointer. 

It is essential to remember, as we stressed in the introduction, that the density operator (6.1) is a formal object, which encompasses the statistical properties of 
the outcomes of a {\it large ensemble} $\scriptE$ of runs issued from the initial state $\scriptD(0)=\hat r(0) \otimes \scriptR (0)$, but which has no direct interpretation. 
In order to understand the various features of a measurement, we need not only to describe globally this ensemble, but to account for properties of {\it individual runs}. 
For instance, we need to explain why each individual run provides a well-defined answer, up or down, and why the coefficient $p_\up$ which enters the expression 
(6.1) can be interpreted as Born's probability, that is, as the relative number of individual  runs having provided the result up within the large ensemble 
$\scriptE$ described by (6.1). This question is known as the ``quantum measurement problem'' \footnote{In the literature there exist
various definitions of the measurement problem. We follow Lalo\"e in ref. \citen{Laloe}.}: Can we make theoretical statements about individual 
quantum measurements, in spite of the irreducibly probabilistic nature of quantum mechanics which deals only with ensembles of runs?

In fact, what we have derived dynamically within the statistical formulation of quantum mechanics is only the global expression (6.1), whereas we would like 
to know whether its two parts have {\it separately} a physical meaning. At first sight, this question looks innocuous. It is tempting to assume that the ensemble 
$\scriptE$ described by (6.1) is the union of two subensembles, with relative sizes $p_\up$ and $p_\down$, described by the states 
$\scriptD_\up =\hat r_\up \otimes \scriptR_\Up$  and $\scriptD_\down = \hat r_\down \otimes \scriptR_\Down$, respectively. 
All the runs in the first subensemble would then be characterised by a value up of the pointer, and correlatively by a spin S in the collapsed state $\up\rangle$. 
However this intuitive statement is fallacious due to a specific {\it quantum ambiguity}, as we now show.

As an illustration, consider first a large set of coins, thrown at random. It is correct to state that this set can be split into two subsets, with coins on the heads 
and tails sides, respectively. Going from random bits to random $q$-bits, consider now a large set of non-polarised spins. By analogy, we might believe in the 
existence of two subsets of spins, pointing in the $s_z=+1$ and $s_z= -1$ directions, respectively. However, we are not allowed to make such an intuitive statement. 
Indeed, we might as well have believed in the existence of two subsets, pointing in the $s_x=+1$ and $s_x=-1$ directions, respectively. Then there would exist 
individual spins pointing simultaneously in two orthogonal directions, which is absurd. Whereas we can ascertain, for the ordinary probability distribution of an 
ensemble of coins, that observing an individual coin will provide a well-defined result, head or tails, our uncertainty remains complete as regard individual spins 
characterised by the quantum distribution of their ensemble. Due to such an ambiguity, which arises from the matrix nature of quantum states, we cannot give a 
meaning, in terms of subensembles, to the separate terms of a decomposition of a mixed density operator. 
This forbids us to make any statement about individual systems in the absence of further information.

The same ambiguity prevails for the measurement model that we are considering. Although the decomposition (6.1) of $\scriptD(\tf)$ as a sum of two terms is 
suggestive, and although a naive interpretation of each term seems to provide the expected result, an {\it infinity of other decompositions exist}, which are 
mathematically allowed, and of which {\it none has a priori a physical meaning}. Our sole determination of this expression is not sufficient to provide an interpretation 
of each term of the decomposition (6.1), and hence to justify, as we wish, the so-called postulates of ideal measurements. 
At this stage, the measurement problem remains open. We have to rely on further arguments for its solution,
{ while our only hope can lie in properties of the apparatus.}

\subsection{The strategy}

\renewcommand{\split}{{\rm split}}
\newcommand{\Csub}{{C{\rm sub}}}

Starting from some time $t_\split$ at which the final state (6.1) has already been reached, we consider {\it all possible decompositions} 
into two terms, 

\BEQ
\label{DDk}
\scriptD= k \scriptD_\sub + (1-k) \scriptD_\Csub                         
\EEQ
of the density operator found above, where $0<k<1$ and where $\scriptD_\sub$ and where $\scriptD_\Csub$ have the mathematical properties of density 
operators (hermiticity, normalisation and non-negativity). 
The above quantum ambiguity does not entitle us to ascribe separately a physical meaning to each 
of the two terms of (6.2) -- and in particular not  to each associated with the two terms of (6.1). 
We are not allowed to regard $\scriptD_\sub$ as a density operator of some real subensemble of $\scriptE$.
However, if, conversely, the full ensemble $\scriptE$ of real runs of the measurement described 
by $\scriptD$ is split into a real subensemble $\scriptE_\sub$ of runs and its complement $\scriptE_\Csub$, each of these must be described by {\it genuine 
density operators} $\scriptD_\sub$ and $\scriptD_\Csub$ that satisfy (6.2) at the time $t_\split$ and that are later on governed by the Hamiltonian $\hat H$.
\footnote{Here it is essential to realize that subensembles are also ensembles, thus satisfying the same evolution though with different initial conditions,
while the linearity of the Liouville-von Neumann equation allows the split up (\ref{DDk}) of $\scriptD$ in separate terms.} 

Although we cannot identify whether an operator $\scriptD_\sub$ issued from a decomposition (6.2) describes the state of S+A for some physical subensemble 
$\scriptE_\sub$, or whether it is only an element of a mathematical identity, we will take it as an initial condition at the time $t_\split$ and solve the equations 
of motion for $\scriptD_\sub(t)$ at subsequent times. This step can again be treated, at least formally, as a process of quantum statistical mechanics; 
its ideas and outcome are presented in \S~6.3.

It turns out that, for a suitable choice of the Hamiltonian of the apparatus, {\it any operator} $\scriptD_\sub(t)$ issued from a decomposition (6.2) of (6.1) tends, 
over a short time, to 

\BEQ
\scriptD_\sub(t) \mapsto \scriptD_\sub(\tf) =\sum_i q_i \hat r_i \otimes \scriptR_i,    
\EEQ
which has the same form as (6.1) except for the values of the weights $q_\up\ge0$ and $q_\down =1- q_\up\ge0$. 
The relaxation time is sufficiently short so that this form is attained at the time $\tf$ determined in Section 5. 
The operators (6.3) are the only {\it dynamically stable} ones.

We have stressed that the operators $\scriptD_\sub(t)$ have not necessarily a physical meaning, but that {\it their class encompasses any physical density operator} 
describing some subset $\scriptE_\sub$ of runs. Since all candidates for such physical density operators reach the form (6.3) at the time $\tf$, we are ascertained 
that {\it the state of} S+A associated with {\it any real subensemble} $\scriptE_\sub$ of runs {\it relaxes} as shown in \S~6.3 and {\it ends up in the form} (6.3). 

The collection of all subensembles $\scriptE_\sub$ of $\scriptE$ possesses the following {\it hierarchic structure}. When two disjoint subensembles 
$\scriptE_\sub^{(1)}$  and $\scriptE_\sub^{(2)}$ merge into a new subensemble $\scriptE_\sub$, the corresponding numbers of runs 
$\scriptN$ and weights $q_i$ ($i=\,\up$ or $\down$) satisfy the standard addition rule 

\BEQ
	\scriptN q_i = \scriptN^{(1)} q_i^{(1)} + \scriptN^{(2)} q_i^{(2)}, \qquad       
	\scriptN = \scriptN^{(1)}+ \scriptN^{(2)}.
\EEQ
Thus, one can prove within the framework of quantum statistical dynamics, not only that the state of S+A describing the full set $\scriptE$ of runs is 
expressed by (6.1), but also that the states describing all of its physical subsets $\scriptE_\sub$ have the form (6.3), where the weights $q_i$ are related
 to one another by the hierarchic structure (6.4). In the minimalist formulation of quantum mechanics which deals only with statistical ensembles, 
 this is the most detailed result that can be obtained about the ideal Curie--Weiss measurement process. 
 An extrapolation is necessary to draw conclusions about individual systems, as will be discussed in \S~6.4.  

\subsection{ Subensemble relaxation}

We consider here the evolution for $t\ge t_\split$ of an operator $\scriptD_\sub(t)$, defined at an initial time $t_\split$ through some mathematical decomposition 
 (6.2) of the density operator (6.1), already reached at the time $t_\split$ for the full ensemble of runs. We have seen in Section 5 that, during the last stage of 
 the registration, S and A can be decoupled. Indeed, each of the two terms of the final state (6.1) of S+A is factorised, so that (6.1) describes a thermodynamic 
 equilibrium in which S and A are correlated only through the equality between the signs of $s_z$ and of the magnetisation of A. After decoupling of S and A, 
 the evolution of $\scriptD_\sub(t)$ is governed by the Hamiltonian $\hat H_{\rm A}$ {\it of the apparatus alone}, and the above correlation will be preserved 
 within $\scriptD_\sub (t)$ at all times $t\ge t_\split$. 
 
 We first show that the initial condition $\scriptD_\sub(t_\split$), although undetermined, must satisfy constraints imposed by the form of the equations 
 (6.1) and (6.2) from which it is issued. As the apparatus is macroscopic, we can represent $\scriptR_\Up$  (or $\scriptR_\Down$) as a 
 {\it microcanonical equilibrium state} characterised by the order parameter $+m_{\rm F}$ (or $-m_{\rm F}$). 
 In the Hilbert space of A, we denote as $|i, \eta\rangle$ (with $i=\Up$ or $\Down$) a basis for the microstates that underlie each microcanonical state 
 $\scriptR_i^\mu$, where the energy and magnetization are taken as constants.
   We then have, with a superscript $\mu$ denoting that the equilibrium state of eq. (6.3)
is now taken in the microcanonical ensemble,

 \BEQ
		\scriptR_i ^\mu= \frac{1}{G}\sum_\eta |i, \eta\rangle\langle i, \eta|,                 
 \EEQ
where $G$ is the large number of values taken by the index $\eta$. \footnote{ In some models one may now disregard the bath, so that
$\eta$ denotes states of the magnet M (see the random matrix model of section 11.2.3 of ref. \citen{ABNOpus}); in general  models it denotes states of M+B}
 Denoting by $ |i, \eta\rangle$ (with $i=\up$ or $i=\down$) the two states 
 $s_z=+1$ or $s_z=-1$ of S, we see that the density matrix (6.1) is diagonal in the Hilbert subspace of S+A spanned by the correlated kets
 $ |i\rangle |i, \eta\rangle$ and that it has no element in the complementary subspace. The {\it non negativity of the two terms} of (6.2) then implies that the ref. \citen{ABNOpus}
 latter property must also be satisfied by the operator $\scriptD_\sub(t_\split)$, which has therefore the form

\BEQ	
 \scriptD_\sub(t_\split) = \sum_{i,i',\eta,\eta'} |i\rangle |i, \eta\rangle K(i,\eta;i',\eta') \langle i'|  \langle i',\eta'|
 \EEQ
The matrix $K$ is Hermitean, non negative  and has unit trace. 
 
 Let us now turn to the Hamiltonian $\hat H_{\rm A}$ that governs the subsequent evolution of $\scriptD_\sub(t)$. We assume here that it contains small terms 
 which produce {\it transitions among the microstates} $|\Up,\eta\rangle $, which have nearly the same energy and nearly the same magnetisation -- likewise 
 among the microstates $|\Down,\eta\rangle $. Although these terms are small, they are very efficient because they practically conserve the energy. 
 Being small, their occurrence does not affect the derivations of \S~4 and \S~5, and conversely the present ``{\it quantum collisional process}''
  is governed solely by 
 the rapid transitions between the kets  $ |i\rangle |i, \eta\rangle$ having the same $i$ but different $\eta$.
 \footnote{ Again, a good apparatus must satisfy the proper requirements for this aspect of the dynamics. 
 Ref.  \citen{ABNOpus}  discusses that it is realistic to assume that apparatuses satisfy them in practice.}
 
 Such a dynamics keeps the form of (6.6) unchanged but modifies the matrix $K$. For a large apparatus, it produces an irreversible process which 
 generalises the {\it microcanonical relaxation} to an intricate situation involving two different microcanonical states. It has been worked out in ref. \citen{ABNOpus} (Section 12);
 the result is the following. Over a short delay, all the matrix elements with $i\neq i'$ (that is, the combinations $\up\down$ and $\down\up$) of 
 $K(i, \eta; i', \eta',t)$ tend to $0$. Over the same delay, its elements $\up \up$ with $\eta\neq \eta'$ also tend to $0$, while the diagonal elements 
 $\up \up$ with $\eta=\eta'$ all tend to one another, their sum remaining constant -- likewise for its elements $\down \down$. 
 Hence, using (6.5), we find that $\scriptD_\sub(t)$ rapidly tends to

 \BEQ 
 \scriptD_\sub(t) \mapsto \sum_i q_i \hat r_i \otimes \scriptR_i^\mu,          
\EEQ
where $\hat r_i = |i\rangle \langle i|$ and $q_i=\sum _\eta K(i, \eta; i,\eta)$. This relaxation holds for any mathematically allowed decomposition 
 (6.2) of (6.1), and in particular for any physical decomposition associated with the splitting of the ensemble of runs of the measurement into subensembles.

\subsection{Emergence of classicality}

 It remains to solve the quantum measurement problem, that is, to understand how we can make statements about individual runs of the process, 
 although quantum theory, in its minimalist statistical formulation, deals only with ensembles. We have already succeeded to determine, for ideal Curie--Weiss 
 measurements treated within this theoretical framework, the expressions (6.3) and (6.4) which embody the strongest possible results about the final states of 
 S+A for {\it arbitrary subensembles} of runs.  
 
 In order to extrapolate this result to the {\it individual runs} which constitute these subensembles, we note that the common form (6.3) of the states $\scriptD_\sub$ 
 and the hierarchic structure (6.4) of the weights are exactly the same as in ordinary probability theory. On the one hand, the difficulties arising from the 
 quantum ambiguity have been overcome owing to a dynamical property, the subensemble relaxation, which produced the stable final states (6.3). 
 On the other hand, the relation (6.4) satisfied by the weights $q_i$ is one of the axioms that define classical probabilities as {\it frequencies of occurrence} 
 of individual events \cite{vMises}.  It is therefore natural to interpret each coefficient $q_i$ associated with a given subset of runs as the proportion of runs 
 of this subset that have yielded the result $i$. In particular, for the full ensemble $\scriptE$, we recover {\it Born's rule}: We had found above $p_\up$ only 
 as a weight that occurred in the decomposition (6.1) of $\scriptD(\tf)$; we can now interpret it as a classical probability, defined as the relative frequency 
 of occurrence of $+m_{\rm F}$ in all the individual runs of $\scriptE$. 

We are then led to {\it interpret}  $\hat r_\up \otimes \scriptR_\Up$ as the density operator associated with the subset for which $q_\up=1$, $q_\down=0$
-- { it is here where {\it interpretation} enters our approach}. 
{ With now having a homogeneous (pure) subensemble at hand,  we can associate this density operator {\it with any individual run} of this subset. }
 Thus, contrary to the first stages of the measurement process, the truncation and the registration, the so-called  ``{\it collapse}'' is not a physical process. 
 It appears merely as a subsequent {\it updating of the density operator} which results from
the selection of a subensemble, {made possible by the effectively vanishing of the off-diagonal terms of the density operator of the full system.}
 
Here again, the apparatus plays a major r\^ole. It is only the observation in a given run of its indication $+m_{\rm F}$ which allows us to predict, 
 owing to the correlations between S and A,  that this run constitutes a preparation of S in the state $|\hspace{-1mm}\up\rangle$.  The emergence in a measurement 
 process of classical concepts, uniqueness of the outcome for an individual event, classical probabilities, classical correlations between S and A, 
 relies on the {\it macroscopic size of the apparatus.}

\renewcommand{\thesection}{\arabic{section}}
\section{An attempt to simultaneously measure non-commuting variables}
\setcounter{equation}{0}\setcounter{figure}{0}
\renewcommand{\thesection}{\arabic{section}.}
\label{section.3}
Textbooks in quantum mechanics (artificially) describe measurements as an instantaneous process, which rules out the possibility of even \emph{trying} to simultaneously measure two non-commuting observables.
Nevertheless, in the Curie-Weiss model, the measurement is described as a physical interaction between the measured system and the apparatus. 
An interesting scenario appears then if one lets the measured system interact with two such apparatuses
\emph{simultaneously}, each of which is attempting to measure a different spin component\cite{ABNOpus}. 

At this point one may argue that this process is meant to fail. Indeed, even if both apparatuses would yield results for their respective measurements, 
it is clear that a quantum state can not have two well definite values for two non-commuting observables (the two different spin components). However, the point of
this section is precisely to find out in which sense this process differs from an ideal measurement, and to give a good interpretation of the obtained results.

In order to set the problem in technical terms, let us consider a general spin state
\begin{equation}
\hat{\rho}(0)=\frac{1}{2}\left\{\mathbb{I}+\langle  \hat{\bf s}(0) \rangle \cdot\hat{\bf s} \right\}.
\label{rho0extwo}
\end{equation}
It will simultaneously interact with two apparatuses A and A$'$, which attempt to measure $\hat{s}_z$ and $\hat{s}_x$, respectively.
By reading the pointers of A and A$'$, we aim to achieve some information about \emph{both} $\langle \hat{s}_z (0) \rangle$ 
and $\langle \hat{s}_x (0) \rangle$ in every run of the experiment. As in any measurement in quantum mechanics, many runs of 
the experiment will be needed to know $\langle \hat{s}_z (0) \rangle$ and $\langle \hat{s}_x (0) \rangle$ with good precision.

Finally, notice that if we were to measure $\hat{s}_z$ and $\hat{s}_x$ sequentially, then the second measurement would be completely
uninformative. For instance, starting form the general state (\ref{rho0extwo}), after measuring  $\hat{s}_z$ the state is 
$\frac{1}{2}\left\{\mathbb{I}+\langle  \hat{s}_z(0) \rangle \hat{s}_z \right\}$, which has no memory about $\langle \hat{s}_x (0)\rangle$. 

\subsection{The Hamiltonian}
We extend the Curie--Weiss model by adding a new apparatus A$'$ made up of a magnet M$'$ and a bath B$'$, with parameters $ J', g', N'...$. 
The total Hamiltonian is then given by $ \hat{H}_{\rm T}=\hat{H}_{\rm SA}+\hat{H}_{\rm SA'}+\hat{H}_{\rm A}+\hat{H}_{\rm A'}$,
with $\hat{H}_{\rm SA}=-Ng\hat{m} \hat{s}_z $ and $\hat{H}_{\rm SA'}=-N'g' \hat{m}' \hat{s}_x$; so each component of the spin is interacting with a different apparatus. 
The internal Hamiltonians $H_{\rm A}$, $H_{\rm A'}$ can be found from (\ref{HA}). Although the apparatuses are not necessarily identical, we
assume them to be similar, i.e.,  $N,J,g,\gamma$ are of the same order of $N',J',g',\gamma'$ respectively.

It will turn out to be very useful to define a direction ${\bf u}$ where the interacting Hamiltonian is diagonal, that is:
\begin{equation}
H_{\rm SAA'} =\hat{H}_{\rm SA}+\hat{H}_{\rm SA'}=\frac{\hbar}{2} w(\hat{m},\hat{m}') \hat{s}_{\bf u} (\hat{m},\hat{m}')
\label{HintTwoApp}
\end{equation}
with $\hat{s}_{\bf u}(m,m')={\bf u}(m,m')\cdot {\bf \hat{s}}$, and 
\begin{eqnarray}
 {\bf u}(m,m')=\frac{2Ngm}{\hbar w} \hat {\bf z} + \frac{2N'g'm'}{\hbar w} \hat {\bf x}
\label{definitionu}
\\
w(m,m')=\frac{2}{\hbar}\sqrt{(Ngm)^2+(N'g'm')^2} 
\label{definitionww}
\end{eqnarray}
Therefore, effectively the spin acts on both apparatuses as a global field $w$ in the direction $\bf{u}$. 
Finally, let us define a direction ${\bf v}$ perpendicular to ${\bf u}$ and $y$, 
\begin{equation}
 \hat{s}_{\bf v}={\bf v}(m,m')\cdot {\bf \hat{s}}=u_z \hat{s}_x - u_x \hat{s}_z.
\end{equation}

\subsection{The state}
The joint state of $\rm{S}+\rm{M}+\rm{M'}$ will be denoted by $\hat{D}(\hat{m},\hat{m}',t)$, and it can be characterized as:
\begin{equation}
 \hat D(\hat{m},\hat{m}',t)=\frac{1}{2G(\hat m)G(\hat m')}\left[ P(\hat m,\hat m',t)+{\bf C}(\hat m,\hat
m',t)\cdot\hat{\bf s}\right].
\label{charstatetwonon}
\end{equation}
In order to interpret this description, consider 

\begin{eqnarray}
 \tr \left\{\delta_{\hat{m},m} \delta_{\hat{m}',m'}  \hat{D} \right\}=P(m,m',t),\hspace{7mm}
 \tr \left\{\delta_{\hat{m},m} \delta_{\hat{m}',m'}\hat{s}_i  \hat{D}\right\}=C_i(m,m',t). 
\end{eqnarray}
where $\delta_{\hat{m},m}$ is a projector on the subspace with magnetization $m$. Therefore, $P(m,m',t)$ is the joint probability distribution 
of the magnetization of the apparatuses and 
$C_i(m,m',t)$, with $i=x,y,z$ or $i=x,u,v$; brings information about the correlations between $\hat{s}_i$ and the apparatuses.

Initially, the system and the apparatuses are uncorrelated, thus being in a product state $\hat{r}(0)\otimes \hat{R}_{\rm M}(0) \otimes \hat{R}_{\rm M'}(0)$
with  $\hat{R}_{\rm M}(0)$ given in (\ref{purePM}). $P_{\rm M}(m)$, given in (\ref{Pm0}), is the probability distribution associated to $\hat{R}_{\rm M}(0)$, 
and the initial state of the correlators is $C_i(0)=\langle \hat{s}_i (0)\rangle P_{\rm M}(m) P_{\rm M}(m')$.

\subsection{Disappearance of the off-diagonal terms}

In the Curie-Weiss model, truncation, or the disappearance of the off-diagonal terms, was shown to be a dephasing effect due to the interacting Hamiltonian. 
Let us thus focus only on the 
action of $H_{\rm SAA'}$, as defined in (\ref{HintTwoApp}), and disregard the other terms of the total Hamiltonian. 
Obviously then the ${\bf u}$-component of the spin is preserved in time, analogously to 
$\hat{s}_z$ for the one apparatus case. On the other hand, by inserting the Ansatz (\ref{charstatetwonon}) into the Liouville-von Neumann equation of motion we find
\begin{equation}
 i  \frac{\partial {\bf C} \cdot \hat{{\bf s}} }{\partial t}=-\frac{1}{2}\left[w \hat{s}_{\bf u}, {\bf C}\cdot \bf{\hat{s}} \right]
\end{equation}
where we also projected onto subspaces with given magnetizations. Using the commuting properties of the Pauli matrices these equations can be readily solved, yielding:
\begin{eqnarray}
P(t)=P(0) \nonumber\\
C_u(t)=C_u(0) \nonumber\\
C_y(t)=C_y(0) \cos(wt) \nonumber\\
C_v(t)=C_v(0) \sin(wt) 
\label{solution2apwithoutbath}
\end{eqnarray}
which shows how the correlators $C_y$ and $C_v$ rapidly rotate because of the external field $w$. 
This situation should be compared with the precessing of the spins in the magnet for the case of one apparatus, see (\ref{rhoneq0}), 
which lead to the decay of the off-diagonal terms (\ref{slong}). The same mechanism is responsible now for the fast decay of 
$\langle \hat{s}_y  \rangle$ and $\langle \hat{s}_v  \rangle$. Furthermore,  the bath-induced decoherence at later times will only increase this effect, yielding
the actual suppression of the correlators $C_y$ and $C_v$ \cite{ABN10,ThesisMarti}.

Therefore, truncation will now occur in the $\bf{u}$ direction. Notice however that $\bf{u}$ is a function of $m$ and $m'$, which in turn will evolve in time as the registration takes place. Therefore, the preferred basis is not fixed, but it keeps changing during the measurement; and the collapse basis will depend on each particular run of the process 
(i.e, on the final values of $m$ and $m'$). This is a signature of the non-ideality of the considered measurement.

\subsection{Registration}
During the registration the magnets are expected to reach ferromagnetic states due to the combined effect of the spin system and the baths. 
This takes place in a longer time scale than the truncation, and it can be described by solving the equations of motion for $P(m,m',t)$ and $C_i(m,m',t)$ including 
the terms arising from the baths. The corresponding equations become notably complex, particularly because $P(m,m',t)$ becomes coupled to all $C_i$; and we refer the 
reader to refs. 
\citen{ABN10,ThesisMarti}
for a detailed analysis of the dynamics. Here instead we will focus our attention on the final state. Since it is an equilibrium state,
much can be said about its characteristics by studying the free energy function.

Notice from (\ref{HintTwoApp}) that the action of the spin on the magnets can be seen as an external field $w$, thus the joint free energy function for the both 
magnets can be written as
\begin{equation}
\mathcal{F}(m,m')=\frac{\hbar}{2}w+F(m)+F(m')
\label{FE2app}
\end{equation}
where $F(m)$ is the free energy of one apparatus in absence of interactions, as given in (\ref{F=}) with $h=0$. In order to find the local stable points where
 the states of the magnets are expected to evolve to, the student can find the local minima of (\ref{FE2app}). Initially setting $w=0$ and $T<0.496J$, 
 one can find a local minima around $(m,m')=(0,0)$; 
four local minima at (0,$\pm m_{\rm F}$) and ($\pm m_{\rm F} , 0$); and four global minimima  at ($\pm m_{\rm F}$,$\pm m_{\rm F}$). 
The paramagnetic state is the initial state for the magnets, which is metastable. On the other hand, if the final state is centered at (0,$\pm m_{\rm F})$ or ($\pm m_{\rm F}$,0),   
then only one of the magnets has achieved registration; whereas if  if is in one the global minima at ($\pm m_{\rm F},\pm m_{\rm F}$), then \emph{both} of them have.   
Finally, one can find the minimum coupling necessary to allow for a rapid transition between the paramagnetic and the ferromagnetic states, i.e., the minimum $g, g'$ 
so that the  free energy barriers disappear.

\subsection{The final state and its interpretation}

We are interested in the final probability distribution of $P(m,m',\tf)$, from which we can extract information about the measured observables $\hat{s}_x$ and $\hat{s}_z$. 
Our study of the free energy function shows that the most stable points are found in $(m,m')=(\pm m_F, \pm m_F)$, and for a sufficiently large coupling we expect the final magnetization of the magnets to evolve towards such points.
These four points are associated with the four possible outcomes of the measurement: $(s_z=\pm \frac{\hbar}{2},s_x=\pm \frac{\hbar}{2})$. The final state thus has the form
\begin{equation}
 P(m,m',\tf)=\sum_{\epsilon =\pm 1} \sum_{\epsilon'=\pm 1} \mathcal{P}_{\epsilon \epsilon' } \delta_{m,\epsilon m_F} \delta_{m',\epsilon' m_F'}
\label{form2measurements}
\end{equation}
where $\delta_{m,x}$ represents a narrow (normalized) peak at $m=x$. $\mathcal{P}_{\epsilon \epsilon'}$, 
which are the weights of each peak, represent the probabilities of getting one of the 4 possible outcomes.

Let us discuss the dependence of the weights $\mathcal{P}_{\epsilon \epsilon'}$ on the initial conditions of S. It has been argued that the correlators 
$C_v$ and $C_y$ disappear due to a dephasing effect together with a decoherence effect at later times.
Therefore only  $C_u(m,m',0)$ contributes, which is a linear combination of $\langle \hat{s}_x(0) \rangle$ and $\langle \hat{s}_z(0) \rangle$. Since the equations of motion 
are linear, the final result for $P(m,m',t)$ (and $C_u$) will still be a linear combination of $\langle \hat{s}_x(0) \rangle$ and $\langle \hat{s}_z(0) \rangle$. On the other 
hand, if $\langle \hat{s}_x(0) \rangle=\langle \hat{s}_z(0) \rangle=0$, then we have $\mathcal{P}_{\epsilon \epsilon'}=1/4$ due to the symmetry $m \leftrightarrow -m$ and $m' \leftrightarrow -m'$. Putting everything together, we can write the general form:

\begin{equation}
 \mathcal{P}_{\epsilon \epsilon'}=\frac{1}{4}[\,1+\epsilon \lambda \langle \hat{s}_z (0) \rangle+\epsilon' \lambda' \langle \hat{s}_x (0) \rangle\,]
\label{Peedeff}
\end{equation}
where $\epsilon,\epsilon'=\pm 1$; and  $\lambda$, $\lambda'$ are the proportionality factors. We term such factors the \emph{efficiency} factors. 

Consider now a particular case where the tested spin is initiall pointing at $+z$, i.e.,  $\langle s_x(0) \rangle=0$ and $\langle s_z(0) \rangle=1$. Then, the probability that A, 
the apparatus measuring $\hat{s}_z$,  ends up pointing at $+m$ is $\mathcal{P}_{++}+\mathcal{P}_{+-}=\frac{1}{2}(1+\lambda)$; whereas there is a probability 
$\mathcal{P}_{-+}+\mathcal{P}_{--}=\frac{1}{2}(1-\lambda)$ to
end up at $-m$, thus yielding a \emph{wrong} indication. Indeed, according to Born rule if $\langle s_z(0) \rangle=1$ then a device measuring $\hat{s}_z$ will always yield the
same outcome, whereas in the current case there is a probability $\frac{1}{2}(1-\lambda)$ of failure. Finally, notice that it must hold $\lambda \in [0,1]$ and similarly it can 
be shown $\lambda' \in [0,1]$.

Since $\mathcal{P}_{\epsilon \epsilon'}$ must be non-negative for any initial state of S, and because (\ref{Peedeff}) has the form 
$\frac{1}{4}(1+{\bf a} \cdot {\bf \hat{s}})$ with $|\bf{a}|\leq 1$, we reach the condition:
\begin{equation}
 \lambda^2+\lambda'^2 \leq 1
\label{conditionslambdalin}
\end{equation}
Therefore, we can already say that both measurements can not be ideal. In the case of two identical apparatuses, such a condition yields:
 $\lambda \leq \frac{1}{\sqrt{2}}$. For example, if $\lambda=\lambda'=1/\sqrt{2}$, starting with a spin pointing in the z direction, $\langle \hat{s}_z(0) \rangle=1$, 
 there is a probability 
of $(1-1/\sqrt{2})/2\approx 0.15$ to read the result $-\hbar/2$ in the apparatus measuring $\hat{s}_z$. Nevertheless, how much information can we extract 
from the results of the apparatuses?

Notice that relation (\ref{Peedeff}) can be inverted:
\begin{eqnarray*}
 \langle \hat{s}_z(0) \rangle= \frac{1}{\lambda}(\mathcal{P}_{++}+\mathcal{P}_{+-}-\mathcal{P}_{-+}-\mathcal{P}_{--}) \nonumber\\
 \langle \hat{s}_x(0) \rangle= \frac{1}{\lambda'}(\mathcal{P}_{++}-\mathcal{P}_{+-}+\mathcal{P}_{-+}-\mathcal{P}_{--})
\label{sxandszexpectvalues}
\end{eqnarray*}
as long as $\lambda$ and $\lambda'$ do not vanish \footnote{In ref. \citen{ThesisMarti} it is shown how $\lambda, \lambda'$ do not vanish and can take values close to $1/\pi$.}. 
Therefore, by counting the different results $\{++, +-, -+ , --\}$ of the experiment we can obtain the weights $P_{ij}$ ($i,j=\pm$) and 
thus $\langle \hat{s}_x (0) \rangle$ \emph{and} $\langle \hat{s}_z(0)\rangle$ with arbitrary precision. 
The fact that we need many runs of the experiment to determine the measured observables $\hat{s}_z$ and $\hat{s}_z$ is a feature of any measurement in quantum mechanics. 
In conclusion, although the process is not ideal (the apparatuses can yield false indications), it is completely informative.

\renewcommand{\thesection}{\arabic{section}}
\section{ General conclusions}
\setcounter{equation}{0}\setcounter{figure}{0}

	The very interpretation, conceptually essential, of quantum mechanics requires an understanding of quantum measurements, experiments which give 
us access to the microscopic reality through macroscopic observations. In a theoretical approach, measurements should be treated as dynamical processes 
for which the tested system and the apparatus are coupled. Since the apparatus is macroscopic, and since the elucidation of the problems related to 
measurements requires an analysis of time scales, we must resort to non equilibrium quantum statistical mechanics.

	This programme has been achieved above for two models, the Curie--Weiss model for ideal measurements (Sections 2--6), and a modified model 
which exhibits the possibility of drawing information about two non commuting observables of S through a large set of runs of non ideal measurements (Section 7). 
It turns out that the questions to be solved pertain to the physics of the apparatus rather than to the physics of the system itself, whether we consider the diagonal 
or the off-diagonal contributions to a density matrix. It is the specific properties of the apparatus and of its coupling with the system which ensure that an experiment 
can be regarded as a measurement providing faithful information about this system.

\begin{table}[ht]
\tbl{The steps of ideal quantum measurements}
{\begin{tabular}{@{}ccccc@{}} \toprule
Descriptive level & full ensemble & full ensemble     & subensembles & individual systems  \\
\colrule
Process  & truncation & registration & relaxation & reduction  \\ \\
Mechanism(s)               & dephasing & phase transition &decoherence & selection  \\     
               & decoherence & energy dump & &  \\     \\
Approach &Q stat mech & Q stat mech &Q stat mech & interpretation\\     
\botrule
\end{tabular}
}
\begin{tabnote}
 For the full ensemble the initial state is the one to be measured, 
       for the subensembles the initial conditions are unknown but constrained by  positivity.
       The so-called ``reduction of the state" or ``collapse of the wave function" is the result of selection of measurement outcomes.
\end{tabnote}
\label{ra_tbl1}
\end{table}

Our theoretical analysis relies solely on standard quantum statistical mechanics. Through such an approach we can acknowledge the emergence of qualitatively 
new phenomena when passing from a microscopic to a macroscopic scale. For instance, in classical statistical mechanics, the irreversibility observed at our scale 
emerges from the microscopic equations of motion that are reversible. This looks paradoxical, but can be explained by the possibility of neglecting correlations 
between a large number of microscopic constituents, which have no physical relevance, and by the inaccessibly large value of recurrence times. 
The irreversibility of quantum measurement processes has the same origin.

Moreover, the same type of approximations, legitimate owing to the macroscopic size of the apparatus and to the properties of its Hamiltonian, allows us to 
understand another kind of emergence. The quantum formalism, which governs objects at the microscopic scale, presents abstract, counterintuitive features 
foreign to our daily experience. In its minimalist formulation, quantum theory deals with statistical ensembles, wave functions or density operators are not reducible 
to ordinary probability distributions; quantities like ``quantum correlations'' cannot be regarded as ordinary probabilistic correlations since they violate 
Bell's inequalities.\footnote{See ref. \citen{TheoBellFoP}  for the opinion that Bell inequality violation implies only that quantum mechanics works, 
without any statement about presence or absence of local realism.}
The quantum theoretical analysis of measurement processes allows us to grasp the emergence of a classical description of their outcome and of classical concepts, 
in apparent contradiction with the underlying quantum concepts (Section 6). In particular the possibility of assigning ordinary probabilities to individual events 
through observation of the apparatus provides a solution to the so called measurement problem.  

We thus conclude that our analysis of ideal quantum measurements involves three steps: 
study of the dynamics of the full ensemble of runs (including truncation and registration), 
study of the final evolution of arbitrary subensembles, 
and inference towards individual systems. See table 1.

We advocate the statistical formulation for the teaching of quantum theory, since it works for our
discussion of ideal measurements where an interpretation of the ``quantum probabilities'' emerges.
The concept of state is simple to grasp by being in spirit close to classical statistical physics.
States described by wave functions should be regarded only as special cases,
since pure and mixed states both describe ensembles.
Non intuitive features of quantum mechanics remain concentrated in the non commutation
of the observables representing the physical quantities.

\section*{Acknowledgments}
The authors thank the students at the Advanced School on Quantum Foundations and Open Quantum Systems  in Joa\~o Pessoa, Brazil,
2012, for their enthusiastic participation and critical remarks, which helped to deepen this presentation.

 \section*{Appendices}

\setcounter{section}{0}
\renewcommand{\thesection}{\Alph{section}}
\section{The phonon bath}

\setcounter{equation}{0}
\setcounter{figure}{0}
\renewcommand{\thesection}{\Alph{section}.}
\label{AppendixA}


The interaction between the magnet and the bath, which drives the apparatus to equilibrium, is taken
as a standard spin-boson Hamiltonian \cite{petr,Weiss,Gardiner}
\begin{equation}
\mytext{\textcurrency ham4\textcurrency \qquad}
\hat{H}_{{\rm M}{\rm B}
}=\sqrt{\gamma}\sum_{n=1}^{N}\left(  \hat{\sigma}_{x}^{\left(  n\right)  }
\hat{B}_{x}^{\left(  n\right)  }+\hat{\sigma}_{y}^{\left(  n\right)  }\hat
{B}_{y}^{\left(  n\right)  }+\hat{\sigma}_{z}^{\left(  n\right)  }\hat{B}
_{z}^{\left(  n\right)  }\right)  \equiv\sqrt{\gamma}\sum_{n=1}^{N}
\sum_{a=x,y,z}\hat{\sigma}_{a}^{\left(  n\right)  }\hat{B}_{a}^{\left(
n\right)  }{ ,} \label{ham4}
\end{equation}
which couples each component $a=x$, $y$, $z$ of each spin $\mathbf{\hat{\sigma}}^{\left(  n\right)}$ with some hermitean linear combination
$\hat{B}_{a}^{\left(  n\right)  }$ of phonon operators. The dimensionless
constant $\gamma\ll1$ characterizes the strength of the thermal coupling
between ${\rm M}$ and ${\rm B}$, which is weak.

For simplicity, we require that the bath acts independently for each spin degree of freedom $n$, $a$. 
(The so-called independent baths approximation.)  This can be achieved ($i$) by
introducing Debye phonon modes labelled by the pair of indices $k$, $l$, 
with eigenfrequencies $\omega_{k}$ depending
only on $k$, so that the bath Hamiltonian is
\begin{equation}
\mytext{\textcurrency Hbath\textcurrency \qquad}
\hat{H}_{{\rm B}}=\sum
_{k,l}\hbar\omega_{k}\hat{b}_{k,l}^{\dagger}\hat{b}_{k,l}{ ,}
\label{Hbath}
\end{equation}
and ($ii$) by assuming that the coefficients $C$ in
\begin{equation}
\mytext{\textcurrency B=\textcurrency \qquad}
\hat{B}_{a}^{\left(  n\right)
}=\sum_{k,l}\left[  C\left(  n,a;k,l\right)  \hat{b}_{k,l}+C^{\ast}\left(
n,a;k,l\right)  \hat{b}_{k,l}^{\dagger}\right]  \label{B=}
\end{equation}
are such that
\begin{equation}
\mytext{\textcurrency CC\textcurrency \qquad}
\sum_{l}C\left(  n,a;k,l\right)
C^{\ast}\left(  m,b;k,l\right)  =\delta_{n,m}\delta_{a,b}\,c\left(  \omega
_{k}\right)  { .} \label{CC}
\end{equation}
This requires the number of values of the index $l$ to be at least
equal to $3N$. For instance, we may associate with each component
$a$ of each spin $\mathbf{\hat{\sigma}}^{\left(  n\right)  }$ a
different set of phonon modes, labelled by $k$, $n$, $a$,
identifying $l$ as ($n$, $a$), and thus define
$\hat{H}_{{\rm B}}$ and $\hat{B}_{a}^{\left(  n\right)  }$ as
\begin{eqnarray}
\mytext{\textcurrency Hbaths\textcurrency \qquad}
\hat{H}_{{\rm B}}  &
=&\sum_{n=1}^{N}\sum_{a=x,y,z}\sum_{k}\hbar\omega_{k}\hat{b}_{k,a}
^{\dagger\left(  n\right)  }\hat{b}_{k,a}^{\left(  n\right)  }{
,}\label{Hbaths}\\
 \hat{B}_{a}^{\left(n\right)  }&=&\sum_{k}\sqrt{c\left(  \omega_{k}\right)  }\left(  \hat{b}
_{k,a}^{\left(  n\right)  }+\hat{b}_{k,a}^{\dagger\left(  n\right)  }\right)
{ .} \label{B=s}
\end{eqnarray}
We shall see in \S ~\ref{section.3.3.2} that the various choices of the phonon set, of the spectrum (\ref{Hbath}) and of the operators (\ref{B=}) coupled to the spins are
equivalent, in the sense that the joint dynamics of ${\rm S}+{\rm M}$ will depend only on the spectrum $\omega_{k}$ and on the coefficients $c\left(\omega_{k}\right)$.

\renewcommand{\thesection}{\Alph{section}}
\section{Equilibrium state of the bath}
\setcounter{equation}{0}
\setcounter{figure}{0}
\renewcommand{\thesection}{\Alph{section}.}


\vspace{3mm}

\label{section.3.3.2}

At the initial time, the bath is set into equilibrium at the temperature\footnote{We recall that we use units where Boltzmann's constant is equal to one; 
otherwise, $T$ and $\beta=1/T$ should be replaced throughout by  $k_BT$ and $1/k_B T$, respectively.} $T=1/\beta$.
The density operator of the bath, 
\begin{equation}
\mytext{\textcurrency RB0=\textcurrency \qquad}
\hat{R}_{{\rm B}}\left(0\right)  =\frac{1}{Z_{{\rm B}}}e^{-\beta\hat{H}_{{\rm B}}}{\rm  ,}
\label{RB0=}%
\end{equation}
when $\hat{H}_{{\rm B}}$ is given by (\ref{Hbath}), describes the set of
phonons at equilibrium in independent modes.

As usual, the bath will be involved in our problem only through
its \textit{autocorrelation function} in the equilibrium state (\ref{RB0=}),
defined  in the Heisenberg picture by

\BEA 
\mytext{\textcurrency Kt-s}  \mytext{\textcurrency \qquad}
&{\rm tr}_{{\rm B}}\left[  \hat{R}_{{\rm B}}\left(  0\right)
\hat{B}_{a}^{\left(  n\right)  }\left(  t\right)  \hat{B}_{b}^{\left(
p\right)  }\left(  t^{\prime}\right)  \right] =\delta_{n,p}\delta
_{a,b}\,K\left(  t-t^{\prime}\right)  {\rm  ,}\label{Kt-s}\\
\mytext{\textcurrency Bt\textcurrency \qquad}
&\hat{B}_{a}^{\left(  n\right)
}\left(  t\right)    \equiv\hat{U}_{{\rm B}}^{\dagger}\left(  t\right)
\hat{B}_{a}^{\left(  n\right)  }\hat{U}_{{\rm B}}\left(  t\right)  {\rm 
,}\label{Bt}\\
\mytext{\textcurrency UB\textcurrency \qquad}
&\hat{U}_{{\rm B}}\left( t \right) = e^{-i\hat{H}_{{\rm B}}t/\hbar}{\rm  ,} \label{UB}%
\EEA
in terms of the evolution operator $\hat U_{\rm B}(t)$ of ${\rm B}$ alone. The bath operators
(\ref{B=}) have been defined in such a way that the equilibrium expectation
value of $B_{a}^{\left(  n\right)  }\left(  t\right)  $ vanishes for all $a=x,y,z$~\cite{petr,Weiss,Gardiner}. Moreover, the
condition (\ref{CC}) ensures that the equilibrium correlations between
different operators $\hat{B}_{a}^{\left(  n\right)  }\left(  t\right)  $ and
$\hat{B}_{b}^{\left(  p\right)  }\left(  t'\right)  $ vanish,  unless $a=b$ and $n=p$, and that the
autocorrelations for $n=p$, $a=b$ are all the same, thus defining a unique
function $K\left(  t\right)  $ in (\ref{Kt-s}).  We introduce the Fourier transform and its inverse,

\begin{equation}
\mytext{\textcurrency TF\textcurrency \qquad}
\tilde{K}\left(  \omega\right)
=\int_{-\infty}^{+\infty}{\rm d}t\ e^{-i\omega t}K\left(  t\right),\qquad
K(t)=\frac{1}{2\pi}\int_{-\infty}^{+\infty}{\rm d}\omega \,e^{i\omega t}\tilde K\left(  \omega\right)
\label{TF}
\end{equation}
and choose for $\tilde K(\omega) $ the simplest expression having the required properties, namely the quasi-Ohmic form
\cite{caldeira,petr,Weiss,Gardiner}

\begin{equation}
\mytext{\textcurrency K\symbol{94}tilde\textcurrency \qquad}
\tilde{K}\left(
\omega\right)  =\frac{\hbar^{2}}{4}\frac{\omega e^{-\left\vert \omega
\right\vert /\Gamma}}{e^{\beta\hbar\omega}-1}{\rm  .} \label{Ktilde2}
\end{equation}
The temperature dependence accounts for the quantum bosonic nature of the
phonons \cite{petr,Weiss,Gardiner}. The Debye cutoff $\Gamma$ characterizes the largest frequencies of
the bath, and is assumed to be larger than all other frequencies entering our problem.
 The normalization is fixed so as to let the constant $\gamma$
entering (\ref{ham4}) be dimensionless.
Since $\tilde K(\omega)$ is real, it holds that $K(-t)=K^\ast(t)$.

\renewcommand{\thesection}{\Alph{section}}
 \section{Elimination of the bath} 
\label{AppendixA}

  \setcounter{equation}{0}
  \setcounter{figure}{0}
 \renewcommand{\thesection}{\Alph{section}.}

\vspace{3mm}

Taking $\hat H_0=\hat H_{\rm S}+\hat H_{\rm SA}+\hat H_{\rm M}$ and $\hat H_{\rm B}$
as the unperturbed Hamiltonians of S + M and of B, respectively, and denoting by
$\hat U_0=\exp(-i\hat H_0/\hbar)$ and $\hat U_{\rm B}=\exp(-i\hat H_{\rm B}/\hbar)$ the corresponding evolution operators, we consider
the full evolution operator associated with $\hat H=\hat H_0+\hat H_{\rm B}+\hat H_{\rm MB}$ 
in the interaction representation. In general we can expand it to first order in $\sqrt{\gamma}$ as

\begin{equation}
\mytext{\textcurrency UUU\textcurrency \qquad}
\hat{U}_{0}^{\dagger}\left(t\right)  \hat{U}_{{\rm B}}^{\dagger}\left(  t\right)  
e^{-i\hat{H}t/\hbar}\approx\hat{I}-i\hbar^{-1}\int_{0}^{t}{\rm d}t^{\prime}\hat{H}
_{{\rm MB}}\left(  t^{\prime}\right)  +{\cal O}\left(  \gamma\right)
{ ,} \label{UUU}
\end{equation}
where the coupling in the interaction picture is
\begin{equation}
\mytext{\textcurrency HMBt\textcurrency \qquad}
\hat{H}_{{\rm MB}}\left(t\right) =
\hat{U}_{0}^{\dagger}\left(t\right)  \hat{U}_{{\rm B}}^{\dagger}\left(  t\right)  H_{\rm MB}
\hat{U}_{{\rm B}}\left(t\right)  \hat{U}_{0}\left(  t\right)  
 =\sqrt{\gamma}\sum_{n,a}\hat{U}_{0}^{\dagger}\left(  t\right)
\hat{\sigma}_{a}^{\left(  n\right)  }\hat{U}_{0}\left(  t\right)  \hat{B}
_{a}^{\left(  n\right)  }\left(  t\right)  { ,} \label{HMBt}
\end{equation}
with $\hat{B}_{a}^{\left(  n\right)  }\left(  t\right)  $ defined by (\ref{Bt}).

We wish to take the trace over ${\rm B}$ of the exact equation of motion eq. (\ref{vN}) 

\begin{equation}
\mytext{\textcurrency vN\textcurrency \qquad}
i\hbar\frac{{\rm d}
{\hat {\cal D}}}{{\rm d}t}=\left[  \hat{H},{\hat {\cal D}}\right]
{\rm  ,} \label{vN}
\end{equation}
for  $\hat{\cal D}(t)$, so as to generate an equation of motion for the density operator $\hat{D}\left(
t\right)  $ of ${\rm S}+{\rm M}$. In the right-hand side the term
${\rm tr}_{{\rm B}}\left[  \hat{H}_{{\rm B}},{\hat {\cal D}
}\right]  $ vanishes and we are left with
\begin{equation}
\mytext{\textcurrency dD1\textcurrency \qquad}
i\hbar\frac{{\rm d}\hat{D}
}{{\rm d}t}=\left[  \hat{H}_{0},\hat{D}\right]  +{\rm tr}
_{{\rm B}}\left[  \hat{H}_{{\rm MB}},{\hat {\cal D}}\right]  { .}
\label{dD1}
\end{equation}

The last term involves the coupling $\hat{H}_{{\rm MB}}$ both directly and
through the correlations between ${\rm S}+{\rm M}$ and ${\rm B}$
which are created in ${\cal D}\left(  t\right)  $ from the time $0$ to the
time $t$. In order to write (\ref{dD1}) more explicitly, we first exhibit
these correlations. To this aim, we expand ${\cal D}\left(  t\right)  $ in
powers of $\sqrt{\gamma}$ by means of the expansion (\ref{UUU}) of its
evolution operator. This provides, using $\hat U_0(t)=\exp[-i\hat H_0t/\hbar]$, 
\begin{equation}
\mytext{\textcurrency UUDUU\textcurrency \qquad}
\hat{U}_{0}^{\dagger}\left(
t\right)  \hat{U}_{{\rm B}}^{\dagger}\left(  t\right)  {\hat {\cal D}
}\left(  t\right)  \hat{U}_{{\rm B}}\left(  t\right)  \hat{U}_{0}\left(
t\right)  \approx{\hat {\cal D}}\left(  0\right)  -i\hbar^{-1}\left[
\int_{0}^{t}{\rm d}t^{\prime}\hat{H}_{{\rm MB}}\left(  t^{\prime
}\right)  \hspace{-1mm},\hat{D}\left(  0\right)  \hat{R}_{{\rm B}}\left(  0\right)
\right]  +{\cal O}\left(  \gamma\right)  { .} \label{UUDUU}
\end{equation}

Insertion of the expansion (\ref{UUDUU}) into (\ref{dD1}) will allow us to
work out the trace over ${\rm B}$. Through the factor $\hat{R}_{{\rm B}
}\left(  0\right)  $, this trace has the form of an equilibrium
expectation value. As usual, the elimination of the bath variables
will produce memory effects as obvious from (\ref{UUDUU}). We wish
these memory effects to bear only on the bath, so as to have a
short characteristic time. However the initial state which enters
(\ref{UUDUU}) involves not only $\hat {R}_{{\rm B}}\left(
0\right)  $ but also $\hat{D}\left(  0\right)  $, so that a mere
insertion of (\ref{UUDUU}) into (\ref{dD1}) would let $\hat
{D}\left(  t\right)  $ keep an undesirable memory of
$\hat{D}\left(  0\right) $. We solve this difficulty by
re-expressing perturbatively $\hat{D}\left( 0\right)  $ in terms
of $\hat{D}\left(  t\right)  $. To this aim we note that
the trace of (\ref{UUDUU}) over ${\rm B}$ provides
\begin{equation}
\mytext{\textcurrency DtD0\textcurrency \qquad}
U_{0}^{\dagger}\left(  t\right)
\hat{D}\left(  t\right)  \hat{U}_{0}\left(  t\right)  =\hat{D}\left(
0\right)  +{\cal O}\left(  \gamma\right)  { .} \label{DtD0}
\end{equation}
We have used the facts that the expectation value over $\hat{R}_{{\rm B}
}\left(  0\right)  $ of an odd number of operators $\hat{B}_{a}^{\left(
n\right)  }$ vanishes, and that each $\hat{B}_{a}^{\left(  n\right)  }$ is
accompanied in $\hat{H}_{{\rm MA}}$ by a factor $\sqrt{\gamma}$. Hence the
right-hand side of (\ref{DtD0}) as well as that of (\ref{dD1}) are power
series in $\gamma$ rather than in $\sqrt{\gamma}$.

We can now rewrite the right-hand side of (\ref{UUDUU}) in terms of $\hat
{D}\left(  t\right)  $ instead of $\hat{D}\left(  0\right)  $ by means of
inserting (\ref{DtD0}), then insert the resulting expansion of ${\hat {\cal D}}\left(
t\right)  $ in powers of $\sqrt{\gamma}$ into (\ref{dD1}). Noting that the
first term in (\ref{UUDUU}) does not contribute to the trace over ${\rm B}
$, we find
\BEA
&&\frac{{\rm d}\hat{D} }{{\rm d}t}-\frac{1}{i\hbar}\left[  \hat{H}_{0},\hat{D}\right] 
= -\frac{1}{\hbar^{2}}{\rm tr}_{{\rm B}}
\int_{0}^{t}{\rm d}t^{\prime}
\\&\times&
\left[  \hat{H}_{{\rm MB}}(0),\hat{U}_{{\rm B}}(t)\hat{U}_{0}(t)\left[  \hat
{H}_{{\rm MB}}\left(  t^{\prime}\right)  ,\hat{U}_{0}^{\dagger}(t)\hat{D}(t)
\hat{U}_{0}(t)\hat{R}_{{\rm B}}\left(  0\right)  \right]  \hat{U}_{0}
^{\dagger}(t)\hat{U}_{{\rm B}}^{\dagger}(t)\right]  +{\cal O}\left(
\gamma^{2}\right)  { ,} \label{dD2bis} \nn
\EEA
where $\hat{H}_{{\rm MB}}(0)$ is just equal to $\hat{H}_{{\rm MB}}$, see eq. (\ref{dD1}).
Although the effect of the bath is of order $\gamma$, the derivation has required only the first-order term, in $\sqrt{\gamma}$, of
the expansion (\ref{UUDUU}) of ${\cal D}\left(  t\right)  $.

The bath operators $\hat{B}_{a}^{\left(  n\right)  }$ appear through $\hat
{H}_{{\rm MB}}$ and $\hat{H}_{{\rm MB}}\left(  t^{\prime}\right)  $, and
the evaluation of the trace thus involves only the equilibrium autocorrelation
function (\ref{Kt-s}). Using the expressions (\ref{ham4}) and (\ref{HMBt}) for
$\hat{H}_{{\rm MB}}$ and $\hat{H}_{{\rm MB}}\left(  t^{\prime}\right)
$, denoting the memory time $t-t^{\prime}$ as $u$, and introducing the
operators $\hat{\sigma}_{a}^{\left(n\right)  }\left(  u\right) $ 
defined by (\ref{sigmaui}), we finally find the differential equation
(\ref{dD}) for $\hat D(t)$.

Notice that by using (C.6) we have written an equation which self consistently couples the time derivative of $\hat D(t)$ 
to $\hat D(t)$ at the same time, at lowest order in $\gamma$. The method  is akin to the derivation of the renormalization group equation.

In our model, the Hamiltonian commutes with the measured observable $\hat
{s}_{z}$, hence with the projection operators $\hat{\Pi}_{i}$ onto the states
$\left\vert \uparrow\right\rangle $ and $\left\vert \downarrow\right\rangle $
of ${\rm S}$. The equations for the operators $\hat{\Pi}_{i}\hat{D}\hat
{\Pi}_{j}$ are therefore decoupled. We can replace the equation (\ref{dD}) for
$\hat{D}$ in the Hilbert space of ${\rm S}+{\rm M}$ by a set of four
equations for the operators $\hat{R}_{ij}$ defined by (\ref{Rij}) in the Hilbert space of ${\rm M}$.
We shall later see (section 8.2) that this simplification underlies the ideality of the measurement process.

The Hamiltonian $\hat{H}_{0}$ in the space ${\rm S}+{\rm M}$
gives rise to two Hamiltonians $\hat{H}_{\uparrow}$ and
$\hat{H}_{\downarrow}$ in the space ${\rm M}$,\ which according
to (\ref{HSA}) and (\ref{HM=}) are simply two functions of the observable $\hat{m}$, given by%
\begin{equation}
\mytext{\textcurrency Hi\textcurrency \qquad}
\hat{H}_{i}=H_{i}\left(  \hat
{m}\right)  =-gNs_{i}\hat{m}-N\frac{J}{4}\hat{m}^{4}{\rm  ,} \qquad (i=\up,\down)
 \label{Hi}
\end{equation}
with $s_{i}=+1$ (or $-1$) for $i=\uparrow$ (or $\downarrow$).
These Hamiltonians $\hat{H}_{i}$, which describe interacting spins
$\mathbf{\hat {\sigma}}^{\left(  n\right)  }$ in an external field
$gs_{i}$, occur in
(\ref{dD}) both directly and through the operators
\begin{equation}
\mytext{\textcurrency sigmaui\textcurrency \qquad}
\hat{\sigma}_{a}^{\left(
n\right)  }\left(  u,i\right)  =e^{-i\hat{H}_{i}u/\hbar}\hat{\sigma}
_{a}^{\left(  n\right)  }e^{i\hat{H}_{i}u/\hbar}{\rm  ,} \label{sigmaui}
\end{equation}
obtained by projection of (\ref{sigmau}) 

\begin{equation}
\mytext{\textcurrency sigmau\textcurrency \qquad}
\hat{\sigma}_{a}^{\left(n\right)  }\left(  u\right)  \equiv\hat{U}_{0}\left(  t\right)  \hat{U}
_{0}^{\dagger}\left(  t^{\prime}\right)  \hat{\sigma}_{a}^{\left(  n\right)
}\hat{U}_{0}\left(  t^{\prime}\right)  \hat{U}_{0}^{\dagger}\left(  t\right)
=\hat{U}_{0}\left(  u\right)  \hat{\sigma}_{a}^{\left(  n\right)  }\hat{U}
_{0}^{\dagger}\left(  u\right)  \label{sigmau}.
\end{equation}
with $\hat{\Pi}_{i}=|i\rangle\langle i|$ and reduction to the Hilbert space of ${\rm M}$, with $i=\up,\down$.

The equation (\ref{dD}) for $\hat{\cal D}(t)$ which governs the joint dynamics of ${\rm S}+{\rm M}$ 
thus reduces to the four differential equations (\ref{dRij}) in the Hilbert space of ${\rm M}$.



 \end{document}